\documentclass[onecolumn,showpacs,preprintnumbers,amsmath,amssymb, longbibliography]{revtex4}
\usepackage{graphicx}
\usepackage{dcolumn}
\usepackage{bm}
\usepackage{epsfig}
\usepackage{subfigure}
\usepackage{xcolor}
\usepackage{bbold}

\newcommand{\bi}{\begin{itemize}}
\newcommand{\ei}{\end{itemize}}
\newcommand{\be}{\begin{eqnarray}}
\newcommand{\ee}{\end{eqnarray}}
\newcommand{\beq}{\begin{equation}}
\newcommand{\eeq}{\end{equation}}
\newcommand{\dd}{\text{d}}

\newcommand{\bbmatrix}{\left( \begin{array}}
\newcommand{\eematrix}{\end{array} \right)}

\begin{document}

\title{Time-optimal Control of a Dissipative Qubit}

\author{Chungwei Lin$^1$\footnote{clin@merl.com}, Dries Sels$^{2,3}$, Yebin Wang$^1$}
\date{\today}
\affiliation{$^1$Mitsubishi Electric Research Laboratories, 201 Broadway, Cambridge, MA 02139, USA \\ $^2$ Department of physics, Harvard University, Cambridge, MA 02138, USA \\ $^3$ Theory of quantum and complex systems, Universiteit Antwerpen, B-2610 Antwerpen, Belgium}

\begin{abstract}
A formalism based on Pontryagin's maximum principle is applied to determine the time-optimal protocol that drives a general initial state to a target state by a Hamiltonian with limited control, i.e., there is a single control field with bounded amplitude. The coupling between the bath and the qubit is modeled by a Lindblad master equation. 
Dissipation typically drives the system to the maximally mixed state, consequently there generally exists an optimal evolution time beyond which the decoherence prevents the system from getting closer to the target state. 
For some specific dissipation channel, however, the optimal control can keep the system from the maximum entropy state for infinitely long. The conditions under which this specific situation arises are discussed in detail. 
The numerical procedure to construct the time-optimal protocol is described. In particular, the formalism adopted here can efficiently evaluate the time-dependent singular control which turns out to be crucial in controlling either an isolated or a dissipative qubit. 
%For state preparation problems the time-optimal control is either bang-bang or bang-singular-bang in the absence of a bath; the transition being controlled by the total evolution time. 
%%%The optimal evolution time can be infinite or finite depending on the chosen initial and final states, but the fidelity saturates to a finite value at infinity.
%%%%%%%%%%%
\end{abstract}

%\pacs{31.15.A-,71.55.-i,73.20.hb}
\maketitle

%%%%%%%%%%%%%%%%%%%%%%%%%%%%%%%%%%%%%%%%%%%%%%%%%%%%%%%%%%%%%%%%%%%%%%%%%%%%
\section{Introduction}

%Quantum state preparation \cite{PhysRevA.97.062343}
%Single-qubit Landau-Zener (LZ) problem \cite{PhysRevLett.111.260501, PhysRevX.8.031086} 
%Bang-singular-bang \cite{PhysRevX.8.031086, PhysRevA.100.022327} 
%quantum memory \cite{PhysRevLett.114.053602}
%Decoherence-free subspace [?]

Modern quantum technology directly utilizes and manipulates the wave function (including the measurements) to achieve the performance beyond the scope of classical physics. Main applications include quantum computation \cite{Neilson_book, book:Kaye_book, Shor:1997:PAP:264393.264406, Grover:1996:FQM:237814.237866, PhysRevLett.79.325, Peruzzo-2014, Farhi_14, PhysRevX.6.031007}, quantum sensing \cite{PhysRevLett.96.010401, AdvancesQuantumMetrology_2011, PhysRevX.6.031033, PhysRevA.96.040304, PhysRevLett.117.110801, LIGO_2011}, and 
quantum communication \cite{PhysRevLett.69.2881, PhysRevA.61.042302, PhysRevLett.67.661, 144km, RevModPhys.66.481, RevModPhys.77.513, RevModPhys.84.621}.
Reliable and fast quantum state preparation is of crucial importance in most, if not all, of these applications. Whether it is to prepare an initial state for cold-atom quantum simulators~\cite{ahmed19}, trapped ion quantum computing \cite{PhysRevX.8.021012, blatt19} or Nitrogen-vacancy-center quantum sensors~\cite{PhysRevLett.112.047601, Lovchinsky836}, they all need some form of coherent control. A universal approach is to use adiabatic state preparation \cite{PhysRevE.58.5355, Brooke-1999, Santoro-2006, RevModPhys.80.1061, Johnson_2010} by slowly varying external control fields. Its simplicity makes it attractive, but to guarantee adiabaticity one often needs a long evolution time, making it susceptible to decoherence. 

Two strategies exist to speed up this process: (i) shortcuts-to-adiabaticity~\cite{demirplak2003adiabatic, demirplak2005assisted, berry2009transitionless, masuda2009fast, odelin19} and (ii) optimal control theory~\cite{gross07}. The first strategy is based on a recently-proven statement~\cite{PhysRevX.9.011034, PhysRevA.98.043436} that any fast-forward drive can be obtained as a unitary transformation of a counter-diabatic drive. %In this approach, the problem of finding a faster protocol can be decomposed into two seperate problems i.e., that of finding a counter-diabatic protocol, and that of finding a unitary transformation that converts the counter-diabatic Hamiltonian into the original Hamiltonian, with modified time-dependent couplings. 
In this approach, the problem of finding a faster protocol can be decomposed into two separate problems: finding a counter-diabatic protocol; and finding a unitary transformation that converts the counter-diabatic Hamiltonian into the original Hamiltonian, with modified time-dependent couplings. 
The second strategy adopts methods from optimal control theory to find fast driving protocols. In most cases the problem is intractable and one has to resort to numerical methods~\cite{Glaser421, PhysRevX.7.021027, PhysRevX.8.031086}.  However, for problems with only a few degrees of freedom, Pontryagin's maximum principle (PMP) \cite{book:Pontryagin, Sussmann_87_01, book:Luenberger, book:GeometricOptimalControl, PhysRevA.97.062343} can be used to construct the optimal driving protocols. 

Here we restrict our attention to a two-level ``qubit'' system with Landau-Zener (LZ) type Hamiltonian and investigate the effect of system-bath coupling to the optimal control solutions. For the closed system, optimal controls were derived in~\cite{PhysRevLett.111.260501, PhysRevX.8.031086} and at the quantum speed limit they were shown to be of bang-singular-bang type~\cite{PhysRevX.8.031086, PhysRevA.100.022327}. However, real systems are always open and we thus address the following questions:  how robust are these controls to decoherence? and how does the control landscape change by nature of the system-bath coupling? In this paper we gain insight into these  questions by considering a state preparation problem, where the control protocol is designed to steer an initial state to a target state in the shortest time. As a generic dissipation eventually drives the system to its maximum entropy state, for some specific dissipation channel the optimal control finds a path to partially preserve the coherence even when the evolution time goes to infinity. The conditions under which this specific situation arises are discussed in detail. 

The rest of the paper is organized as follows. In Section II we specify the problem and summarize relevant conclusions from classical control theory. The dynamics for the density matrix are introduced to take the dissipations into account. In Section III we consider the optimal control for state preparation problems where the initial and the target state are different. % In Section IV we consider quantum memory problems where the target state is the initial state.
For a special dissipation channel, non-intuitive results are found; the unique aspect of this dissipation will be pointed out and discussed.  A brief conclusion is given in Section VI. 
In Appendices we show a result of numerical optimization and  provide an interesting example (same initial and target states) to support statements made in the main text.

\section{Qubit control as a Landau-Zener Problem} 

Throughout this paper, we describe the qubit control in the context of the Landau-Zener problem. In this section the connection between the qubit control and classical control theory will be provided. We first define the problem by
specifying initial and target qubit states, and the Hamiltonians that can steer the former to the latter. We then cast the qubit-control problem as a time-optimal control problem and summarize the relevant results from PMP. Finally we express the dynamics of the wave function and density matrix dynamics in terms of three real dynamical variables so that the established conclusions from classical control theory can straightforwardly apply. 
As both quantum mechanics and PMP use the term ``Hamiltonian'', to avoid any potential confusions we shall use ``Hamiltonian'' (symbol $H$) in the quantum-mechanical sense; and use ``c-Hamiltonian'' (symbol $\mathcal{H}$) to represent the control-Hamiltonian.

\subsection{Problem statement}
We consider the following single-qubit control problem \cite{PhysRevLett.111.260501, PhysRevX.8.031086}:  
\beq 
\begin{aligned}
H(t; u) & = \sigma_x  + u(t) [\xi \sigma_x + \sigma_z] \\
&\equiv H_0 + u(t) H_d,  \text{ with } |u(t)| \leq 1.
\end{aligned}
\label{eqn:Problem}
\eeq 
In Eq.~\eqref{eqn:Problem}, $\xi$ in $[\xi \sigma_x + \sigma_z] $ is a model parameter whose value will be determined later; the control $u(t)$ is bounded; and $\sigma$'s are Pauli matrices defined as 
\beq 
\sigma_x = \begin{bmatrix} 0&1\\1&0 \end{bmatrix},\,\,
\sigma_y = \begin{bmatrix} 0&-i\\i&0 \end{bmatrix},\,\,
\sigma_z = \begin{bmatrix} 1&0\\0&-1 \end{bmatrix}. 
\nonumber %\label{eqn:Pauli}
\eeq 
The initial and target states are chosen respectively as the ground states of $\sigma_x  + 2 \sigma_z$ and  $\sigma_x  - 2 \sigma_z$, i.e.,
\beq 
\begin{aligned}
|\psi_i \rangle &= \frac{1}{ \sqrt{ 10+4\sqrt{5} }  } \begin{bmatrix} 1\\ -2-\sqrt{5} \end{bmatrix}; \\
|\psi_f \rangle &= \frac{1}{ \sqrt{ 10-4\sqrt{5} }  } \begin{bmatrix} 1 \\ 2-\sqrt{5}  \end{bmatrix}. 
\end{aligned}
\label{eqn:init_final}
\eeq 
We use the same initial and final states chosen in Ref.~\cite{PhysRevLett.111.260501}; the motivation is that these two states should be sufficiently far away from each other to allow for the potentially non-trivial control protocol. Using the Bloch sphere representation where any general state can be represented by three angles (one of them is the overall phase)
\beq 
|\psi (\theta, \phi, \phi_0 ) \rangle 
= e^{i \phi_0} \begin{pmatrix} \cos (\theta/2) \\ e^{i \phi} \sin (\theta/2)  \end{pmatrix}, 
\nonumber
\eeq
we have $\theta_i \approx 0.85 \pi$ and $\theta_f \approx 0.15 \pi$ [see Fig.~\ref{fig:singular_arc}(b) and (d)]. The $\xi$ introduced in Eq.~\eqref{eqn:Problem} determines the ``singular arc'' which will be formally introduced in Section \ref{sec:singular} [see the description following Eq.~\eqref{eqn:[f,g]} for an explicit example].

For the typical time-optimal control problem, one finds the optimal $u^*(t)$ that steers $|\psi_i \rangle$ to $|\psi_f \rangle$ (up to an arbitrary phase)  in the shortest time. %For the memory application, one finds the optimal $u^*(t)$  that keeps $|\psi_i \rangle$ as close to itself as possible during the time evolution. 
For later discussions, Hamiltonians of $|u|=1$ are defined as 
\beq 
\begin{aligned}
H_X &= H_0 - H_d, \\
H_Y &= H_0 + H_d,
\end{aligned}
\label{eqn:H_XY}
\eeq 
i.e., $H_X$ corresponds to $u=-1$ whereas $H_Y$ to $u=+1$. The dissipation effects will be formulated in Section \ref{sec:damping}.

\subsection{Pontryagin's Maximum Principle and the optimality conditions}

The necessary conditions for an optimal solution derived from PMP are discussed in this subsection. To study the quantum system, we consider the control-affine control system, where the dynamics of its ``state variables'' $\pmb{x}$ are described by 
\beq 
 \dot{ \pmb{x} } = \mathbf{f}(\pmb{x}) + u(t) \,\mathbf{g}(\pmb{x}),\,\, \pmb{x} \in R^n, u \in R.
 \label{eqn:control-affine}
\eeq  
$\pmb{x}$ will be referred to as ``dynamical variables'' which can be components of the wave function or the density matrix; $\mathbf{f}$ and $\mathbf{g}$ are smooth vector fields which are functions of $\pmb{x}$. $\mathbf{f}$ is usually referred to as the ``drift'' field as its effect is always present; $\mathbf{g}$ as the ``driving'' field whose strength is controlled by $u(t)$. The admissible range of $u$ is assumed to be bounded by $|u| \leq 1$. Given Eq.~\eqref{eqn:control-affine}, an optimal control $u^*(t)$ minimizes the cost function  
\beq 
\begin{aligned}
J &= \lambda_0 \int_0^{t_f} \dd t + \mathcal{C}(\pmb{x}(t_f) ) \\
&= \lambda_0 t_f  + \mathcal{C}(\pmb{x}(t_f) ),
\end{aligned}
\label{eqn:J}
\eeq  
where $t_f$ is the total evolution time, $\lambda_0$ is a constant, and $\mathcal{C}(\pmb{x}(t_f))$ is a terminal cost function depending only on the values of the dynamical variables at $t_f$. The explicit form of $\mathcal{C}(\pmb{x}(t_f))$ is constructed based on the specific task we would like to accomplish. We only consider the time-invariant problem where $\mathbf{f}$, $\mathbf{g}$, and $\mathcal{C}$ do not depend explicitly on time $t$. The sign of $\lambda_0$  deserves some attentions and becomes important when regarding the evolution time $t_f$ as an optimization variable in Eq.~\eqref{eqn:J}. If $\mathcal{C}(\pmb{x}(t_f) )$ decreases as $t_f$ increases, $\lambda_0$ has to be {\em positive} to allow for a non-trivial optimal $t_f$ (otherwise the optimal $t_f$ is infinity). $\lambda_0 > 0$ corresponds to the conventional time-optimal control problem where one is seeking for the {\em minimum} time to accomplish a certain task. If $\mathcal{C}(\pmb{x}(t_f) )$ increases as $t_f$ increases,   $\lambda_0$ has to be {\em negative} to allow for a non-trivial optimal $t_f$ (otherwise the optimal $t_f$ is zero). $\lambda_0 < 0$ corresponds to finding a {\em maximum} time to achieve a task. The latter case is seldom discussed in classical control theory, but arises naturally in the damped qubit studied here. 
%\textcolor{red}{We stress that the unitary dynamics generally results in an oscillating $\mathcal{C}(\pmb{x}(t_f) )$ (as a function of $t_f$), so the ``optimal time'' can be minimum or maximum time even for the same problem.}

PMP \cite{book:Luenberger} defines a control-Hamiltonian (c-Hamiltonian):
\beq 
\begin{aligned}
\bar{ \mathcal{H} }_c (t)
&= \lambda_0 + \langle \pmb{\lambda}(t), \mathbf{f}(\pmb{x}) \rangle + u(t) \langle \pmb{\lambda}(t), \mathbf{g}  (\pmb{x}) \rangle \\
&\equiv \lambda_0 + \langle \pmb{\lambda}(t), \mathbf{f}(\pmb{x}) \rangle + u(t) \Phi(t) \\
&\equiv \lambda_0 + \mathcal{H}_c (t).
\end{aligned}
\label{eqn:c-Hamil_01}
\eeq 
$\pmb{\lambda}$ is referred to as a set of ``costate'' variables (or the conjugate momentum), which has the same dimension of $\pmb{x}$. $\langle \cdot, \cdot \rangle$ is the inner product introduced for two real-valued vectors. A switching function $\Phi(t) $ is defined as 
\beq 
\Phi(t) = \langle \pmb{\lambda}(t), \mathbf{g}  (\pmb{x}) \rangle, 
\label{eqn:switch_func_01}
\eeq 
which plays the most important role in determining the structure of optimal control. 
Given an optimal solution $(\pmb{x}^*, \pmb{\lambda}^*; u^*)$ to the time-optimal control problem, it has to satisfy the following necessary conditions:
\begin{subequations}
\begin{align} 
& \dot{ \pmb{x} }^*(t) = + \left( \nabla_{\pmb{\lambda}} \mathcal{H}_c\right), \,\,\,\pmb{x}^*(0) \text{ is given.}  
\label{eqn:original_dynamics}
\\
%%%
& \dot{ \pmb{\lambda} }^*(t) = - \left( \nabla_{\pmb{x}} \mathcal{H}_c\right)^T, \,\,\, \pmb{\lambda}^*(t_f) =  \nabla_{\pmb{x}} \mathcal{C} |_{ \pmb{x}^*(t_f) }\label{eqn:costate_dynamics} \\
%%% & \pmb{\lambda}(t_f) =  \nabla_{\pmb{x}} \Psi |_{t=t_f},  \label{eqn:costate_tf} \\
%%% 
& \bar{ \mathcal{H} }_c = 
\lambda_0 + \mathcal{H}_c = \text{ const. } 
\label{eqn:const_c-H} \\ 
& u^*(t) = \begin{cases} +1 & \text{ if } \Phi(t)<0 \\
          -1 & \text{ if } \Phi(t)>0      \\
          \text{undetermined} & \text{ if } \Phi(t)=0 \end{cases}.
\label{eqn:bang_condition}
\end{align}
\label{eqn:necc_cond_state_variables}
\end{subequations}
Eq.~\eqref{eqn:original_dynamics} is equivalent to the dynamics defined in Eq.~\eqref{eqn:control-affine}. Eq.~\eqref{eqn:costate_dynamics} defines the dynamics of costate variables, whose boundary condition is fixed at the final time $t_f$. Eq.~\eqref{eqn:const_c-H} holds for the time-invariant problem. If the final time $t_f$ is not fixed (i.e., $t_f$ is allowed to vary to minimize $\mathcal{C}$), then $\bar{ \mathcal{H} }_c = 0 $. 
Depending the sign of $\lambda_0$ we distinguishes two scenarios for $\mathcal{H}_c$ (instead of $\bar{ \mathcal{H} }_c$): 
%we conclude that $\mathcal{H}_c(t)$ has to be a negative constant for a time-optimal solution. In the following analysis, we focus on $\mathcal{H}_c$ (instead of $\bar{ \mathcal{H} }_c$), and the condition \eqref{eqn:const_c-H} is replaced by 
\beq 
\mathcal{H}_c =   \begin{cases} 
           -|\lambda_0|   \equiv  -1 & \text{ minimum-time solution, } \\
           +|\lambda_0| \equiv   +1 & \text{ maximum-time solution.}
                 \end{cases}
\label{eqn:negative_c-H}
\eeq 
As a function of $t_f$, the terminal cost function is minimized when $\mathcal{H}_c = 0$. Eq.~\eqref{eqn:bang_condition} implies that the optimal control takes the extreme values ($\pm 1$ in this case) when the switching function is nonzero, and is referred to as a bang (B) control. If $\Phi(t) = 0$ over a finite interval of time, the optimal $u^*$ is undetermined from Eq.~\eqref{eqn:bang_condition} and may not take its extreme values; this is referred to as a singular (S) control. The procedure to determine the singular $u(t)$ for systems having two and three real dynamical variables will be described in Section \ref{sec:singular} [Eqs.~\eqref{eqn:optimal_control} and \eqref{eqn:singular_3d}].

It is worth noting that the switching function [Eq.~\eqref{eqn:switch_func_01}] corresponds to the gradient of the terminal cost function and can be used in any gradient-based optimization algorithms  \cite{PhysRevX.9.011034, KHANEJA2005296}. 
When the exact optimal control is not known, the optimality conditions listed in Eqs.\eqref{eqn:necc_cond_state_variables} provide an formalism to quantify the quality of any numerically obtained protocol. 
In the Appendix \ref{subsec:numeric} we give an example to show that the gradient-based method can capture the singular control despite the optimal control has a vanishing $\Phi(t)=0$ over a finite interval of time.

\subsection{Evaluation of singular control \label{sec:singular}}

The density matrix for a qubit  involves three real-valued dynamical variables, and the general formalism to determine the singular control for two and three dynamical variables is now provided. 
A singular arc corresponds to a state trajectory where the switching function vanishes over a {\em finite} interval of time, i.e.,  $\Phi(t) = \dot{\Phi}(t) = \ddot{\Phi}(t) = ... = \Phi^{(n)}(t) = 0$ along the singular arc. The switching function and its first and second time derivatives are given by 
\beq 
\begin{aligned}
\Phi(t) &= \langle \pmb{\lambda}, \mathbf{g} \rangle, \\
\dot{\Phi}(t) &= \langle \pmb{\lambda}, [\mathbf{f}, \mathbf{g}] \rangle, \\
\ddot{\Phi}(t) &= \langle \pmb{\lambda}, \left[\mathbf{f}, [\mathbf{f}, \mathbf{g}] \right] \rangle
+ u \langle \pmb{\lambda}, \left[\mathbf{g}, [\mathbf{f}, \mathbf{g}] \right] \rangle.
\end{aligned}
\eeq 
Here the commutator between two vector fields generates a new vector field given by 
$
\mathbf{h}^i = \left( [\mathbf{f}, \mathbf{g}]\right)^i \equiv  
\langle\mathbf{f},  (\nabla \mathbf{g}^i) \rangle - 
\langle \mathbf{g}, (\nabla \mathbf{f}^i)\rangle, 
$ 
with $\mathbf{f}^i$ being $i$th component of the vector field $\mathbf{f}$.

To determine the singular control of two dynamical-variable systems, we only need $\Phi(t) = \dot{\Phi}(t) = 0$ \cite{PhysRevA.100.022327}.
Over the singular arc, Eq.~\eqref{eqn:negative_c-H} imposes 
$ \langle \pmb{\lambda}, \mathbf{f} \rangle \equiv + 1$ or -1 depending on the problems, and the following derivation assumes  $ \langle \pmb{\lambda}, \mathbf{f} \rangle = - 1$.
 Expanding $ [\mathbf{f}, \mathbf{g}] = \alpha \mathbf{f} + \beta \mathbf{g}$,
we get 
$
\dot{\Phi} = \langle \pmb{\lambda},  [\mathbf{f}, \mathbf{g}] \rangle
= \langle \pmb{\lambda}, \alpha \mathbf{f} + \beta \mathbf{g}\rangle = -\alpha. 
$
$\dot{\Phi}(t) = -\alpha=0$ defines a singular arc and a state trajectory. To stay along $\alpha=0$, the control has to satisfy 
\beq 
\begin{aligned}
& L_{\mathbf{f} + u\mathbf{g}} \alpha = 0 
= \frac{1+u}{2} L_\mathbf{Y} \alpha + 
\frac{1-u}{2} L_\mathbf{X} \alpha  \\
\Rightarrow \,\,& 
u_\text{sing} = \frac{ L_\mathbf{X} \alpha+ L_\mathbf{Y} \alpha  }{ L_\mathbf{X} \alpha - L_\mathbf{Y} \alpha }.
\end{aligned}
\label{eqn:optimal_control}
\eeq 
Here $L_\mathbf{Z} \alpha \equiv \langle \mathbf{Z}, \nabla \alpha \rangle$ is the Lie derivative of $\alpha$ with respect to the vector field $\mathbf{Z}$ -- it is the change of $\alpha$ along the direction defined by $\mathbf{Z}$ \cite{book:GeometricOptimalControl}. 
The admissible control $|u|\leq 1$ requires that  $L_\mathbf{X} \alpha$ and $L_\mathbf{Y} \alpha $ have opposite signs. 
%More details can be found in Ref.~\cite{PhysRevA.100.022327}.

For systems composed of three dynamical variables, we need $\Phi(t) = \dot{\Phi}(t) = \ddot{\Phi}(t) = 0$ to determine values of the singular control. Using $\mathbf{f}$, $\mathbf{g}$, and $[\mathbf{f}, \mathbf{g}]$ as a complete basis, we expand
\beq 
\begin{aligned}
\left[ \mathbf{f}, [\mathbf{f}, \mathbf{g}] \right] &= 
\alpha_1 \mathbf{f} + \alpha_2 \mathbf{g} + \alpha_3 [\mathbf{f}, \mathbf{g}], \\
%%%%%%%
\left[ \mathbf{g}, [\mathbf{f}, \mathbf{g}] \right] &= 
\beta_1 \mathbf{f} + \beta_2 \mathbf{g} + \beta_3 [\mathbf{f}, \mathbf{g}]
\end{aligned} 
\label{eqn:3D_expansion}
\eeq 
to get $\langle \pmb{\lambda}, \left[ \mathbf{f}, [\mathbf{f}, \mathbf{g}] \right] \rangle = -\alpha_1$ and $\langle \pmb{\lambda}, \left[ \mathbf{g}, [\mathbf{f}, \mathbf{g}] \right] \rangle = -\beta_1$ along the singular arc. $\ddot{\Phi}(t) = 0$ determines the value of singular control
\beq 
u_\text{sing} = - \frac{ \langle \pmb{\lambda}, \left[ \mathbf{f}, [\mathbf{f}, \mathbf{g}] \right] \rangle }{ \langle \pmb{\lambda}, \left[ \mathbf{g}, [\mathbf{f}, \mathbf{g}] \right] \rangle } 
= - \frac{\alpha_1}{\beta_1}.
\label{eqn:singular_3d}
\eeq 
Eq.~\eqref{eqn:optimal_control} and \eqref{eqn:singular_3d} respectively determine the state-dependent singular control for 2-dimensional and 3-dimensional cases. The formalism involving commutators (also referred to as the Lie bracket) is termed as ``geometric control technique'' \cite{book:GeometricOptimalControl}. The key usefulness of Eq.~\eqref{eqn:optimal_control} and \eqref{eqn:singular_3d} lies in the fact that the value of singular control at a given $\pmb{x}$ can be computed using only $\mathbf{f}(\pmb{x})$ and $\mathbf{g}(\pmb{x})$ without knowing the entire trajectory.
If the obtained singular control is not admissible one takes the closest bang value. In the numerical simulations, we assume an  optimal $u(t)$ composed of a few bang and singular segments, and use the Nelder-Mead optimization algorithm to determine the switching times. 
The obtained solutions are checked against the necessary conditions given in Eqs.~\eqref{eqn:necc_cond_state_variables}.

\subsection{Application to the single-qubit wave function}

We briefly recapitulate how to express the switching function and c-Hamiltonian in terms of the wave function, more details can be found in Ref. \cite{PhysRevA.100.022327}.
The dynamics of the system is governed by the Schr\"odinger's equation:
\beq 
\begin{aligned}
i \frac{\dd}{\dd t} | \Psi(t) \rangle &= 
\left[ H_0 + u(t) \, H_d \right] | \Psi(t) \rangle,% \\ \text{with }
%| \Psi(t) \rangle  
%&= \begin{bmatrix} \Psi_0 \\ \Psi_1 \end{bmatrix}
%= \begin{bmatrix} \Psi_{0,R} + i \Psi_{0,I} \\ 
%\Psi_{1,R} + i \Psi_{1,I} \end{bmatrix}.
\end{aligned}
\label{eqn:dynamics_grover_explicit}
\eeq 
The initial and target states are given in Eq.~\eqref{eqn:init_final}. To make the final state as close to $|\psi_f \rangle$ as possible, the terminal cost function can be chosen as 
\beq 
\mathcal{C}( \Psi(t_f) ) = -\frac{1}{2} | \langle \psi_f | \Psi (t_f) \rangle |^2.
\label{eqn:cost_Grover}
\eeq 
Using the property that $H_0$ and $H_d$ are real-valued, we can express the c-Hamiltonian and switching function as 
\beq 
\begin{aligned}
\mathcal{H}_{c} &= \text{Im} \langle \Pi(t) | 
\left[ H_0 + u(t) \, H_d \right] | \Psi(t) \rangle, \\
\Phi(t) &=\text{Im} \langle \Pi(t) |  H_d  | \Psi(t) \rangle,
\end{aligned} 
\label{eqn:quantum_Hoc_switch}
\eeq 
where $| \Pi(t) \rangle $ denotes the costate or conjugate momentum to $| \Psi(t) \rangle$. Applying Eq.~\eqref{eqn:costate_dynamics}, one can derive that the dynamics of $|\Pi(t) \rangle$ are governed by the same Schr\"odinger's equation, with the boundary condition given at $t_f$ \cite{PhysRevA.97.062343, PhysRevX.7.021027}: 
\beq 
\begin{aligned}
i \frac{\dd}{\dd t} | \Pi(t) \rangle = 
\left[ H_0 + u(t) \, H_d \right] | \Pi(t) \rangle, \\
\text{ with }|\Pi(t_f) \rangle = -  | \psi_f \rangle \langle \psi_f | \Psi(t_f) \rangle.
\end{aligned}
\label{eqn:Pi_dynamics}
\eeq 
Note that the costate at time $t_f$ is the target state rescaled by its overlap with state at $t=t_f$. 
%The boundary condition $|\Pi(t_f) \rangle$ uses the fact that $|\psi_f \rangle$ has no imaginary part.
%The condition \eqref{eqn:bang_condition} still holds, with the switching function computed using Eq.~\eqref{eqn:quantum_Hoc_switch}. 
%The subscript `Q' in Eqs.~\eqref{eqn:quantum_Hoc_switch} indicates `quantum' and will be dropped from now on. 

\subsection{Application to the single-qubit density matrix \label{sec:damping}} 

%In general, one can of course 
To apply PMP to control open quantum systems, one needs to generalize the previous discussion from unitary dynamics on quantum states to dissipative dynamics on density matrices. This can be done on a formal level, but we will restrict the discussion to the case of a two-level system described by a Markovian master equation. 
Defining $\pmb{\sigma} = (\sigma_x, \sigma_y, \sigma_z)$, $\mathbb{1}$ the identity matrix, $\mathbf{h} = (h_x, h_y, h_z)$, $\pmb{\rho} = (\rho_x, \rho_y, \rho_z)$, a general single qubit Hamiltonian $H$ and density matrix $\rho$ can be parametrized by 
\beq
\begin{aligned}
H &=   \mathbf{h} \cdot \pmb{\sigma}, \\
\rho &= \frac{ \mathbb{1} }{2}  + \frac{1}{2} \pmb{\rho} \cdot \pmb{\sigma}.
\end{aligned}
\eeq 
%Here $\mathbb{1}$ is the identity matrix.
%$\text{Tr}[ \rho ] = 1$ fixes the coefficient of $e$ to be 1/2; all eigenvalues of $\rho$ are non-negative imposes $|\pmb{\rho}| \equiv \sqrt{\rho_x^2 + \rho_y^2 + \rho_z^2} \leq 1$. $|\pmb{\rho}|$ is a measure of the coherence, and $|\pmb{\rho}|=0$ (or equivalently $\rho = e/2$) corresponds to the maximum-entropy state of the qubit system. 
%$\pmb{\rho}$ is a three-dimensional vector that completely specifies the qubit density matrix $\rho$, and we shall use $\pmb{\rho}$ as dynamical variable to describe the qubit dynamics.retention

The dynamics of the system is taken to be governed by the ``Lindblad'' master equation \cite{book:open_quantum}: 
\beq  
\begin{aligned}
\dot{\rho}(t) = \mathcal{L}[ \rho ] &= -i[ H, \rho(t)] 
+\sum_{k=x,y,z} \gamma_k \left[ L_k \rho(t) L^\dagger_k - \frac{1}{2}
 L_k^\dagger L_k \rho(t)  -\frac{1}{2} 
 \rho(t) L^\dagger_k L_k
\right] \\
&\equiv -i[ H, \rho(t)] 
+\sum_{k=x,y,z} \gamma_k \mathcal{D}_k (\rho(t) ),
\end{aligned}
\label{eqn:Lindblad_equation}
\eeq 
where $\gamma_k$ is a positive number specifying the dissipation strength and $L_k$'s are the associated Lindblad operators. Using $\pmb{\rho}$ as dynamical variables, direct calculations give 
%To formulate the problem using the vector $\pmb{\rho}$, we define three ``projection'' operators to describe these three dissipations, i.e. 
\begin{subequations}
\begin{align}
\gamma_x D_{\sigma_x}(\rho) \rightarrow -2 \gamma_x \mathbf{P}_{x}\cdot \pmb{\rho}
= - \Gamma_x \, \text{Diag}(0,1,1) \cdot \pmb{\rho} ,
\label{eqn:sigma_x}\\
%%%
\gamma_x D_{\sigma_y}(\rho) \rightarrow -2 \gamma_y \mathbf{P}_{y}\cdot \pmb{\rho}
= - \Gamma_y \, \text{Diag} (1,0,1) \cdot \pmb{\rho} ,
\label{eqn:sigma_y} \\
%%%
\gamma_x D_{\sigma_z}(\rho) \rightarrow -2 \gamma_z \mathbf{P}_{z}\cdot \pmb{\rho}
= - \Gamma_z \, \text{Diag} (1,1,0) \cdot \pmb{\rho} .
\label{eqn:sigma_z}
\end{align}
\label{eqn:damping}
\end{subequations}
Here $\Gamma_i = 2 \gamma_i$ and $\mathbf{P}_{x,y,z}$ is an operator ($3 \times 3$ matrix) that annihilates the $x,y,z$ component. Because all off-diagonal components of $\mathbf{P}_{x,y,z}$ are zero, only the diagonal components are given in Eqs.~\eqref{eqn:damping}.  Each equation in Eqs.~\eqref{eqn:damping} describes one dissipation channel. For the $\sigma_x$/$\sigma_y$/$\sigma_z$ dissipation channel, only $\rho_x$/$\rho_y$/$\rho_z$ survives in the steady-state solution, the remaining two components will decay to zero.

Using Eq.~\eqref{eqn:Lindblad_equation}, the equation of motion for $\pmb{\rho}$ is
\beq 
\frac{\dd }{\dd t} \pmb{\rho}  = 2 \mathbf{h} \times \pmb{\rho} 
- \Gamma_i \mathbf{P}_i \cdot \pmb{\rho}.
\eeq 
Three costate variables are denoted by $\pmb{\lambda} = (\lambda_x, \lambda_y, \lambda_z)$, and the c-Hamiltonian $\mathcal{H}_{c}$ is
\beq 
\mathcal{H}_{c} = 
\pmb{\lambda} \cdot \left[ 2 \mathbf{h} \times \pmb{\rho} 
- \Gamma_i \mathbf{P}_i \cdot \pmb{\rho} \right].
\label{eqn:c_H_DM}
\eeq 
The equation of motion for $\pmb{\lambda}$ is 
\beq 
\frac{\dd \pmb{\lambda} }{\dd t} = - \frac{\partial \mathcal{H}_{c} }{ \partial \pmb{\rho} } 
=  2 \mathbf{h} \times \pmb{\lambda} + \Gamma_i \mathbf{P}_i \cdot \pmb{\lambda}.
\eeq 
Note that the term that damps $\pmb{\rho}$ becomes the gain for $\pmb{\lambda}$. For $\mathbf{h}(t) = \mathbf{h}_0 + \mathbf{h}_1 u(t)$, the switching function is defined as 
\beq 
\Phi(t) =  2 \pmb{\lambda}(t) \cdot \left[ \mathbf{h}_1 \times \pmb{\rho}(t) \right]. 
\eeq 
For the LZ problem defined in Eq.~\eqref{eqn:Problem}, $\mathbf{h}_0 = \hat{x}$ and $\mathbf{h}_1 = \xi \hat{x} + \hat{z}$.  

The initial state of $| \psi_i \rangle $ corresponds to an initial density matrix $\pmb{\rho} (t=0) = \pmb{\rho}_i = \left(\frac{-1 }{ \sqrt{5}} , 0, \frac{-2 }{\sqrt{5}} \right)$. Similarly the target state of $| \psi_f \rangle $ corresponds to $\pmb{\rho}_f = \left(\frac{-1 }{ \sqrt{5}} , 0, \frac{2 }{\sqrt{5}} \right)$. 
%The specific form of the terminal cost function and boundary conditions for $\pmb{\rho}$ and $\pmb{\lambda}$ are now specified.
%For a general one-qubit state $\langle \Psi (\theta, \phi) | 
%= \begin{pmatrix} \cos \frac{\theta}{2}, & \sin \frac{\theta}{2} e^{-i \phi} \end{pmatrix} $, one gets 
%$ \langle \Psi (\theta, \phi) | \rho |\Psi (\theta, \phi) \rangle 
%&=\frac{1}{2} \langle \Psi (\theta, \phi) | (e + \pmb{\rho} \cdot \pmb{\sigma}) |\Psi (\theta, \phi) \rangle \\
%&= \frac{1}{2} \left[ 1 + \begin{pmatrix} \rho_x & \rho_y & \rho_z \end{pmatrix} 
%\begin{pmatrix} \sin \theta \cos \phi \\ 
%\sin\theta \sin \phi \\ \cos \theta \end{pmatrix} \right] 
%= \frac{1}{2} \left( 1 + \pmb{\rho} \cdot \pmb{n} \right)
%$
%with $\pmb{n}=(\sin \theta \cos \phi, \sin\theta \sin \phi, \cos \theta)$ . %($\theta$ is chosen such that both $\cos \theta \geq 0$ and $\sin \theta \geq 0$). 
For the target state $|\psi_f \rangle$ defined in Eq.~\eqref{eqn:init_final}, the terminal cost function (which we want to minimize) can be chosen as:
%\beq 
%\begin{aligned}
%\mathcal{\tilde{C} } (\pmb{\rho}(t_f) ) =-\frac{1}{2} \left[ 1 + \langle \pmb{\rho}_f , \pmb{\rho} (t_f) \rangle \right]. 
%%\rightarrow - \left[ 1 + \rho_z (t_f) \right] 
%%\left[ \frac{1}{2} \left[ 1 + \rho_z (t_f) \right] - 1 \right]^2 
%% =\frac{1}{4} (\rho_z (t_f)-1)^2.
%\end{aligned}
%\label{eqn:cost_DM}
%\eeq 
%In the simulation we use the negative of $\langle \pmb{\rho}_f , \pmb{\rho} (t_f) \rangle$ as the terminal cost function:
\beq 
\mathcal{C} (\pmb{\rho}(t_f) ) = -\langle \pmb{\rho}_f , \pmb{\rho} (t_f) \rangle, 
\label{eqn:cost_overlap}
\eeq 
whose range is between -1 and 1. The boundary condition of the costate variable  is
\beq 
\pmb{\lambda}(t_f) = \nabla_{\pmb{\rho}(t_f)} \mathcal{C} (t_f) 
= -\pmb{\rho}_f = -\begin{pmatrix} \frac{-1 }{\sqrt{5}}, & 0, &  \frac{2}{\sqrt{5}}  \end{pmatrix}.
\eeq  
The term $\langle \pmb{\rho}_f , \pmb{\rho} (t_f) \rangle $ will be referred to as ``target-state overlap'' in this paper.
Note that for the maximum entropy state where $\pmb{\rho}=0$ (or $\rho = \mathbb{1}/2$), the terminal cost function \eqref{eqn:cost_overlap} is zero. 

\section{Application to the state preparation problem \label{sec:LZ} } 

\subsection{Overview}

In this section we consider the state preparation for dissipative single-qubit systems. Mathematically, the state preparation problems can be mapped to the conventional time-optimal control problem where one tries to find an optimal control that guides the initial state to the target state in the minimum time. 
Without dissipation, the qubit dynamics can be completely described by two real dynamical variables and the detailed analysis is presented in Ref.~\cite{PhysRevA.100.022327}. With dissipation, we naturally expect that there exists an optimal $t_f$ beyond which the target-state overlap $\langle \pmb{\rho}_f , \pmb{\rho} (t_f) \rangle $ can only decrease. We shall show that this intuition is basically true. For the $\sigma_x$ dissipation channel, however, the optimal $t_f$ to maximize  $\langle \pmb{\rho}_f , \pmb{\rho} (t_f) \rangle $ can become infinite and its origin will be discussed.

\begin{figure}[ht]
\begin{center}
\includegraphics[width=0.7\textwidth]{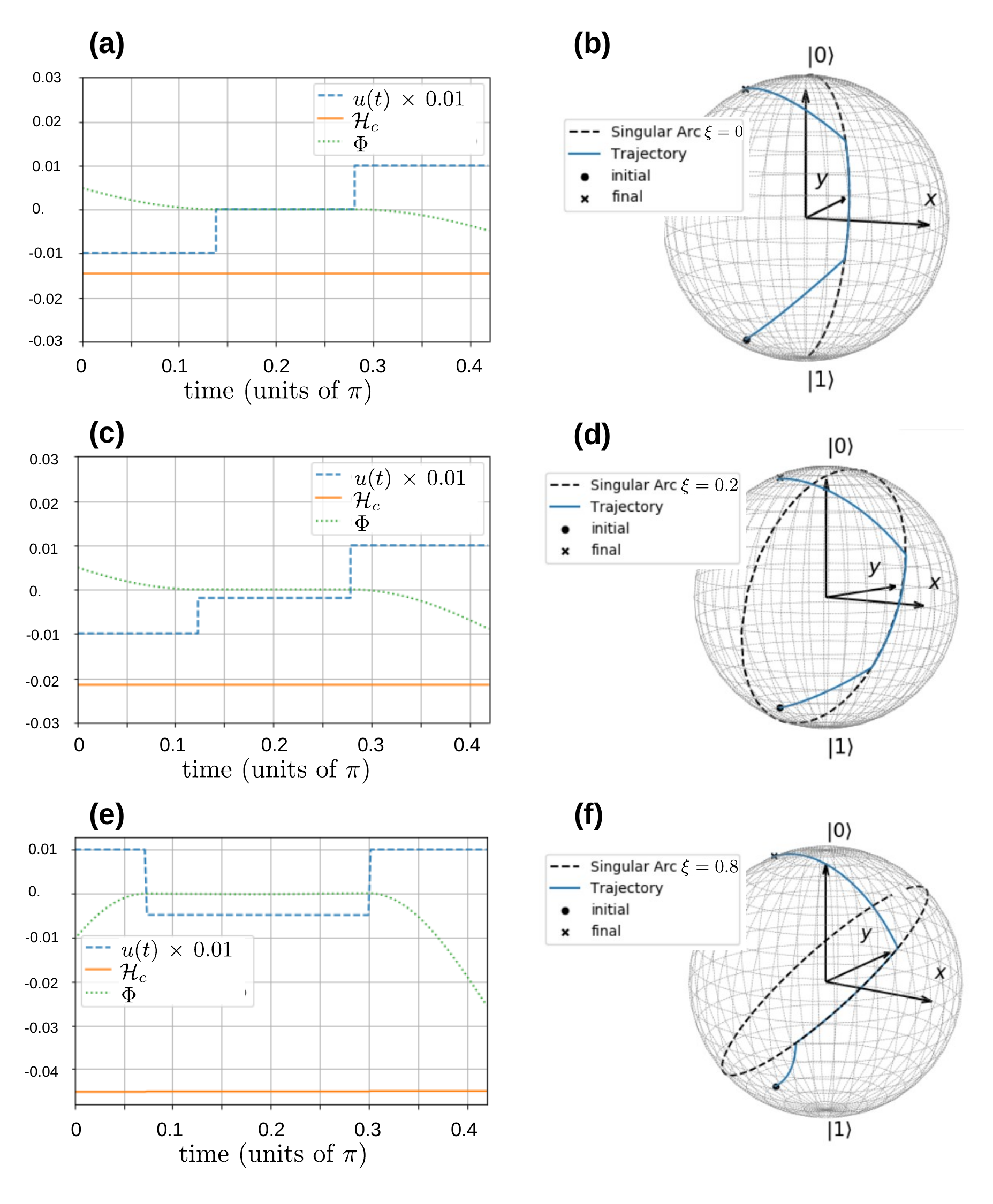}
\caption{The optimal control for the evolution time $t_f=0.42 \pi$. (a) and (b) for $\xi=0$; (c) and (d) for $\xi=0.2$. (e) and (f) for $\xi=0.8$. (a), (c) and (e) show that all necessary conditions are satisfied. Dashed curves: scaled control; solid curves: c-Hamiltonian; dotted curves: switching function. (b), (d) and (f) show the corresponding singular arc (dashed curves) and the optimal trajectory (solid curves) on Bloch sphere. Note that the optimal control goes from XSY to YSY upon increasing $\xi$. 
}
\label{fig:singular_arc}
 \end{center}
\end{figure} 

\subsection{The structure of time-optimal controls -- no dissipation \label{subsec:no_diss}}

To provide a reference for subsequent discussions, we determine the structure the optimal control without dissipation. The detailed formalism, the geometric control technique, is provided in Ref.~\cite{PhysRevA.100.022327} and here we simply use the results. 
We have shown in Ref.~\cite{PhysRevA.100.022327} that without dissipation, the dynamics of a qubit can be described using two real variables $(\theta, \phi)$ and a qubit Hamiltonian corresponds to a two-dimensional vector field defined in the tangent space of $(\theta, \phi)$ manifold. Vector fields corresponding to three Pauli matrices are
\begin{subequations}
\begin{align}
\sigma_z &\rightarrow V_z = 2 \partial_\phi,
\label{eqn:vector_pauli_z}
\\ %%%
\sigma_x &\rightarrow V_x = -2 \sin\phi \, \partial_\theta 
- 2 \cos \phi \cot\theta \,  \partial_\phi ,
\label{eqn:vector_pauli_x}
\\ %%
\sigma_y &\rightarrow V_y = 2 \cos\phi \, \partial_\theta  
-2 \sin \phi  \cot \theta \, \partial_\phi.
\label{eqn:vector_pauli_y}
\end{align}
\label{eqn:vector_pauli}
\end{subequations}
For the LZ problem defined in Eq.~\eqref{eqn:Problem}, we identify $\mathbf{f} \rightarrow V_x$, $\mathbf{g}\rightarrow \xi V_x + V_z$, the commutator $[\mathbf{f},\mathbf{g}]$ is: 
\beq 
%\begin{aligned}
[ \mathbf{f}, \mathbf{g} ] =  2 V_y 
= 4 \begin{bmatrix} \cos \phi \\ -\sin \phi \cot \theta \end{bmatrix}  
%= \alpha \begin{bmatrix} -2 \sin \phi \\ -2\cos \phi \cot \theta \end{bmatrix} 
%+ \beta \begin{bmatrix} -2x \sin \phi \\ 
%-2x \cos \phi \cot \theta + 2 \end{bmatrix}  \\
%%% 
%& = (\alpha + \beta x) \begin{bmatrix} -2 \sin \phi \\ -2\cos \phi \cot \theta \end{bmatrix} 
%+ \beta  \begin{bmatrix} 0 \\ 2 \end{bmatrix}
\equiv \alpha(\theta, \phi) \mathbf{f} + \beta (\theta, \phi) \mathbf{g},
%\end{aligned}
\label{eqn:[f,g]}
\eeq 
where $\alpha(\theta, \phi)$ is found to be $-\frac{2}{ \sin \phi } \left( \cos \phi - \xi \cot \theta \right)$.
The singular arc is defined by $\alpha=0$, i.e., $\xi = \tan \theta \cos \phi$. The parameter $\xi$ thus determines the singular arc, a curve on the surface of Bloch sphere. When $\xi=0$, $\alpha=0$ corresponds to $\phi=\frac{\pi}{2}$ and $\frac{3\pi}{2}$. 
To determine the singular control, we compute  
\beq 
\begin{aligned}
L_{V_z} \alpha &= \frac{4}{\sin^2 \phi} (1 - x \cot \theta \cos \phi) \underset{\alpha=0}{\rightarrow} 4, \\
L_{V_x} \alpha&=\frac{4 \xi}{\sin^2 \theta} + \frac{4 \cos \phi \cot \theta}{\sin^2 \phi} (-1 + \xi \cot \theta \cos \phi) \\
%%%%%%%%%%%%%%%
&\underset{\alpha=0}{\rightarrow}
\frac{4 \xi}{\sin^2 \theta} - 4 \cos \phi \cot \theta=4\xi,
\end{aligned}
\eeq 
from which we can compute $L_{\mathbf{Y}} \alpha = (1+\xi)  L_{V_x} \alpha + L_{V_z} \alpha$ and $L_{\mathbf{X}} \alpha = (1-\xi)  L_{V_x} \alpha - L_{V_z} \alpha$. Substituting into Eq.~\eqref{eqn:optimal_control} we get
%\beq 
%\begin{aligned}
%L_{\mathbf{Y}} \alpha &=  (1+x)  L_{V_x} \alpha + L_{V_z} \alpha
%= 4 \left\{ (1+x) x + 1 \right\} 
%= 4(x^2+x+1)>0, \\
%L_{\mathbf{X}} \alpha &= (1-x)  L_{V_x} \alpha - L_{V_z} \alpha
%= 4 \left\{ (1-x)x- 1 \right\}
%=4(-x^2+x-1)<0. 
%\end{aligned}
%\label{eqn:Lie_Grover}
%\eeq 
%Substituting Eq.~\eqref{eqn:Lie_Grover} into \eqref{eqn:optimal_control}, %we see that $\alpha = 0$ corresponds to a singular control with $u=0$. 
the singular control
\beq 
u_\text{sing} = \frac{ L_{\mathbf{X}} \alpha + L_{\mathbf{Y}} \alpha }{L_{\mathbf{X}} \alpha - L_{\mathbf{Y}} \alpha}
= -\frac{\xi  }{  1+\xi^2} .
\label{eqn:singular_value}
\eeq 
%For the current problem, the optimal control goes from XY to XSY as $t_f$ increases. 
The results for $t_f=0.42 \pi$, $\xi=0, 0.2$ and 0.8 are given in Fig.~\ref{fig:singular_arc}. The trajectories of $(\theta(t), \phi(t))$ can be visualized on a Bloch sphere [Fig.~\ref{fig:singular_arc}(b), (d), and (f)], from which we clearly see that the optimal trajectory and the singular arc overlap over a finite amount of time. Upon increasing $\xi$, the singular arc tilts more (i.e., closer to the equator of the Bloch sphere) and the optimal control changes from XSY [Fig.~\ref{fig:singular_arc} (a) and (c)] to YSY [Fig.~\ref{fig:singular_arc} (e)].

%For $\xi=0$, the equation of motion is reduced to $\dot{\theta}(t) = -2$ [$\phi=\pi/2$ in Eqs.~\eqref{eqn:vector_pauli_y}]. 
It is interesting to consider the case of $\xi=0$ with unbounded $|u(t)|$. As the singular arc is defined by $\phi = \pi/2$, the trajectory under the time-optimal BSB  control, that steers a general initial state $| \psi_{i} \rangle \leftrightarrow (\theta_0, \phi_0)$
to a general target state $| \psi_\text{target} \rangle \leftrightarrow (\theta_1, \phi_1)$,
goes through $(\theta_0, \phi_0) \underset{\text{B}}{\rightarrow}  (\theta_0, \pi/2) \underset{\text{S}}{\rightarrow}  (\theta_1, \pi/2) \underset{\text{B}}{\rightarrow}  (\theta_1, \phi_1)$.
%\beq 
%\begin{pmatrix} \cos (\theta_0/2) \\ e^{i \phi_0} \sin (\theta_0/2)  \end{pmatrix} 
%\underset{\text{B}}{\rightarrow} 
%\begin{pmatrix} \cos (\theta_0/2) \\ i\sin (\theta_0/2)  \end{pmatrix} 
%\underset{\text{S}}{\rightarrow} 
%\begin{pmatrix} \cos (\theta_1/2) \\ i\sin (\theta_1/2)  \end{pmatrix} 
%\underset{\text{B}}{\rightarrow} 
%\begin{pmatrix} \cos (\theta_1/2) \\ e^{i \phi_1} \sin (\theta_1/2 ) \end{pmatrix}.
%\eeq 
B can be X or Y depending on the initial and final states. 
The times of the first and the last bang-control are both infinitesimal; the singular control takes the time $|\theta_0 - \theta_1|/2$ \cite{explanation00}. Expressing $| \psi_{i} \rangle = i_0 | 0 \rangle + i_1 |1 \rangle =  \cos (\theta_0/2) |0\rangle + e^{i \phi_0} \sin (\theta_0/2) |1 \rangle$ and  $| \psi_\text{target} \rangle = t_0 | 0 \rangle + t_1 |1 \rangle = \cos (\theta_1/2) |0\rangle + e^{i \phi_1} \sin (\theta_1/2) |1 \rangle$, the minimum evolution time is given by 
\beq 
\begin{aligned}
T_\text{min} &= \arccos \left(  \cos \frac{\theta_0}{2} \cos \frac{\theta_1}{2} + \sin \frac{\theta_0}{2} \sin \frac{\theta_1}{2} \right) \\ 
&= \arccos( |i_0 t_0| + |i_1 t_1| ) . 
\end{aligned}
\eeq 
This is ``quantum speed limit'' obtained in Ref.~\cite{PhysRevLett.111.260501, PhysRevA.90.032110, PhysRevA.94.023624}.  

The focus of this paper is the dissipative system, and we will use $\xi=0.2$ as the primary example. Results of $\xi=0.8$ will be shown to demonstrate the generality of some nonintuitive behavior found in systems with the $\sigma_x$ dissipation channel. Without dissipation, the minimum time to reach the target state is about 0.44$\pi$ for $\xi=0.2$. %Some results of $\xi=0.8$ will be given in the Appendix to show the generality of our conclusions. 

\begin{figure}[ht]
\begin{center}
\includegraphics[width=0.8\textwidth]{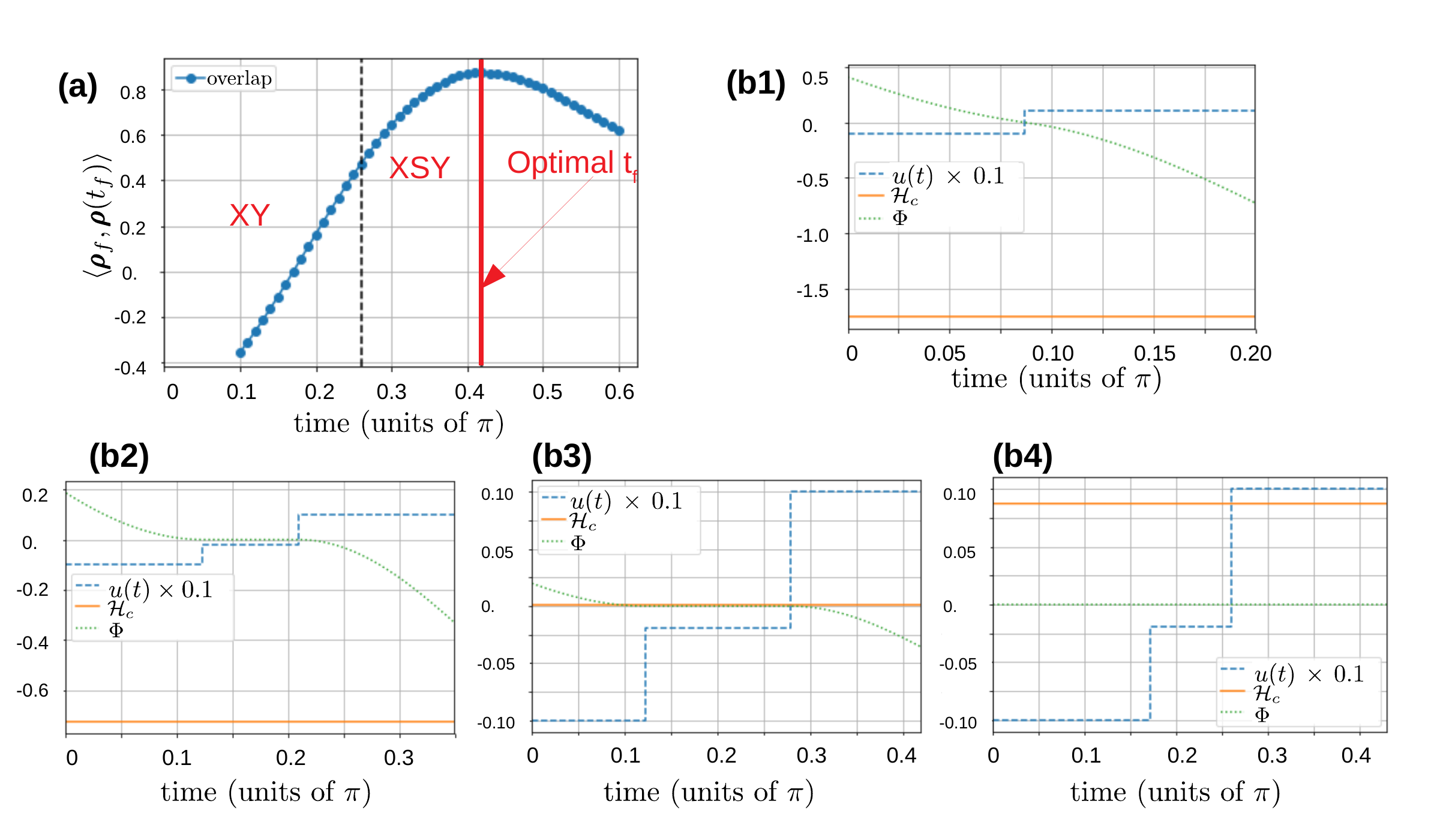}
\caption{$\xi=0.2$ and  $\Gamma = 0.1$ the uniform dissipation described in Eq.~\eqref{eqn:uniform_damp}.  (a) The target-state overlap as a function of the evolution time $t_f$. The vertical dashed (black) line are boundaries of different optimal control structures; corresponding optimal controls are indicated in red. The optimal evolution time is around $0.42 \pi$, indicated by the solid (red) vertical line. (b1)-(b4) The optimal controls at four representative computational times: (b1) $t_f=0.2 \pi$; (b2) $t_f=0.35 \pi$; (b3) $t_f=0.419 \pi$ (optimal $t_f$); (b4) $t_f=0.43 \pi$. Dashed curves: scaled control; solid curves: c-Hamiltonian; dotted curves: switching function.  As the evolution time increases, the optimal control changes from XY (b1) to XSY (b2)-(b4).  For (b1)-(b3), all the necessary conditions are satisfied. (b3) At the optimal evolution time, the c-Hamiltonian is zero. (b4) Beyond the optimal evolution time, the c-Hamiltonian becomes positive.
}
\label{fig:LZ_DampXYZ_x0.2_Gamma0.1}
 \end{center}
\end{figure} 

\subsection{Optimal protocol for the uniform, $\sigma_y$, $\sigma_z$ dissipation channels \label{subsec:uniform_diss}} 

We now take the dissipation into account. Let us first consider the ``uniform''  dissipation (dampings on $\rho_x$, $\rho_y$, $\rho_z$ are identical) where 
\beq 
\frac{\dd}{\dd t} \pmb{\rho} = 2\mathbf{h}(\xi) \times \pmb{\rho} -\Gamma \pmb{\rho}.
\label{eqn:uniform_damp}
\eeq 
We take $\xi=0.2$ and $\Gamma = 0.1$.  Taking the inner product of $\pmb{\rho}$ and Eq.~\eqref{eqn:uniform_damp} gives  
\beq 
\pmb{\rho} \cdot \frac{\dd}{\dd t} \pmb{\rho} 
= \frac{1}{2} \frac{\dd }{\dd t} (|\pmb{\rho}|^2) = -\Gamma |\pmb{\rho}|^2. 
\label{eqn:amplitude_equation}
\eeq 
The amplitude decays exponentially in time: $|\pmb{\rho}(t)|^2 = e^{-2 \Gamma t}$ or $|\pmb{\rho}(t)| = e^{-\Gamma t}$. In this case we expect an optimal evolution time, as $|\pmb{\rho}|$ eventually decays to zero. Because the dynamics of the amplitude is known, one can use $(\theta, \phi)$ as dynamical variables and apply the formalism in Section \ref{subsec:no_diss}. The results are summarized in Fig.~\ref{fig:LZ_DampXYZ_x0.2_Gamma0.1}(a). Upon increasing the evolution time, the optimal control goes from XY to XSY, the same as the closed system. The optimal evolution time is around $0.42\pi$, slightly shorter than 0.44$\pi$ obtained in the closed system. Necessary conditions are checked in Fig.~\ref{fig:LZ_DampXYZ_x0.2_Gamma0.1}(b1)-(b4). We note that the c-Hamiltonian goes from a negative constant to a positive when $t_f$ crosses its optimal value. %We have tested that the XSY protocol applies to other $x$ values. 
Qualitatively similar behaviors are found the for $\sigma_y$ and $\sigma_z$ dissipation channels [see Fig.~\ref{fig:LZ_DM_tscan_DampX_Gamma0p1_xxp2_rhot}(b) for the $\sigma_z$ dissipation channel].% some results of $\sigma_z$ dissipation channel using $\xi=0.2$ and $\Gamma_z = 0.1$ are given in Appendix \ref{App:sigma_z}. 

%%%%%%%%%%%%%%%%%%%%%%%%
%%%%%%%%%%%%%%%%%%%%%%%%%%%%%%%%%%%%%%%
\subsection{Optimal protocol for the $\sigma_x$ dissipation channel \label{subsec:sigma_x}} 

\begin{figure}[ht]
\begin{center}
\includegraphics[width=0.7\textwidth]{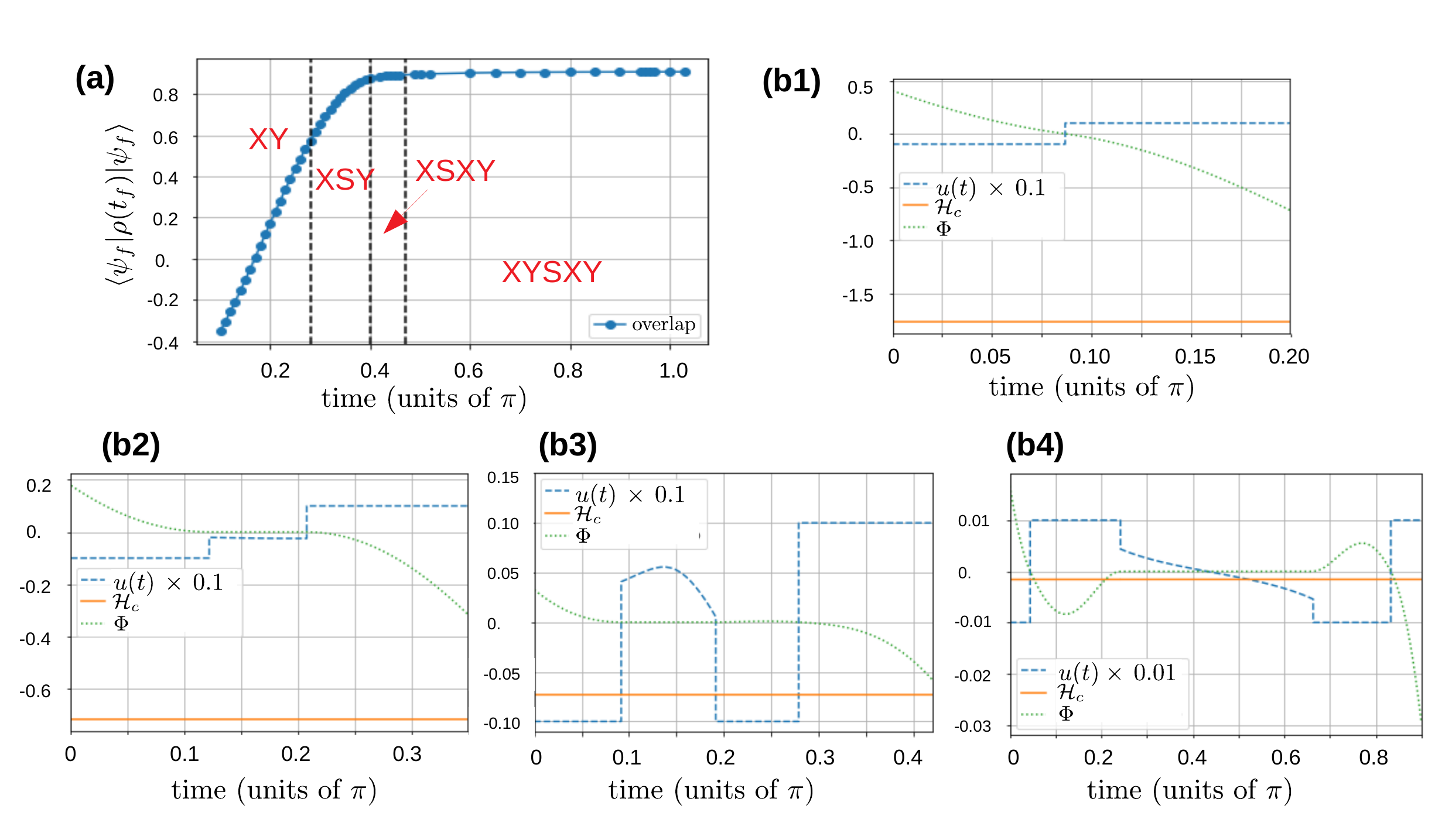}
\caption{ (a) The target-state overlap $\langle \pmb{\rho}_f , \pmb{\rho} (t_f) \rangle $ as a function of evolution time $t_f$. Vertical dashed (black) lines are boundaries of different optimal control structures; corresponding optimal controls are indicated in red. (b1)-(b4) The optimal controls at four representative computational times(b1)-(b4) The optimal controls at four representative computational times: (b1) $t_f=0.2 \pi$; (b2) $t_f=0.35 \pi$; (b3) $t_f=0.42 \pi$; (b4) $t_f=0.90 \pi$. Dashed curves: scaled control; solid curves: c-Hamiltonian; dotted curves: switching function. As the evolution time increases, the optimal control changes from XY (b1) to XSY (b2) to XSXY (b3) to XYSXY (b4). Numerically no finite optimal $t_f$ is found in this case.
}
\label{fig:LZ_DM_tscan_DampX_Gamma0p1_xxp2}
 \end{center}
\end{figure}

The behavior of $\sigma_x$ dissipation channel is qualitatively different from those of uniform, $\sigma_y$, and $\sigma_z$ dissipation channels.  The most important property turns out to be the existence of a one-dimensional null-space of the drift field defined by the $\sigma_x$ dissipation channel. Specifically, the null-space is given by $\pmb{\rho}_c = (\rho_x, 0,0)$ that satisfies 
\beq 
\mathbf{f}(\pmb{\rho}_c) = 
2 \hat{x} \times  \pmb{\rho}_c + \Gamma_x \mathbf{P}_x \cdot \pmb{\rho}_c = 0. 
\label{eqn:rho_c_sigma_x}
\eeq 
For other dissipation channels, $\mathbf{f}(\pmb{\rho}) = 0$ implies $\pmb{\rho}=0$.

%\textcolor{blue}{
The existence of a one-dimensional null-space has direct consequences for the optimality condition in the presence of a singular arc in the optimal protocol. It is {\em a priori} {\em not} clear whether there will be such singular controls, but it appears to be general at least for single qubit problems \cite{PhysRevLett.111.260501, PhysRevX.8.031086, PhysRevA.100.022327}.
Recall that, at the optimal $t_f$, it is required that the c-Hamiltonian vanishes [note that the meaning of optimal $t_f$ depends on the terminal cost function $\mathcal{C}(\pmb{x}(t_f))$, see the discussion below Eq.~\eqref{eqn:negative_c-H}], i.e. at the optimal $t_f$,
\beq 
\mathcal{H}_c(t) = 0 =
\langle \pmb{\lambda} | \mathbf{f} (\pmb{\rho}(t)) \rangle 
+ u^*(t) \langle \pmb{\lambda} | \mathbf{g} (\pmb{\rho}(t)) \rangle.
\eeq 
If the optimal control includes a singular arc, where $\langle \pmb{\lambda} | \mathbf{g} (\pmb{\rho}) \rangle = 0$, then $\mathcal{H}_c(t) = 0$ implies
\beq 
\langle \pmb{\lambda} | \mathbf{f} (\pmb{\rho}) \rangle = 0
\label{eqn:optimal_tf_condition}
\eeq 
along the singular arc. 
Eq.~\eqref{eqn:optimal_tf_condition} is automatically satisfied at $\pmb{\rho} = \pmb{\rho}_c$ because $\mathbf{f}(\pmb{\rho}_c)=0$. However, if $\pmb{\rho}(t)$ indeed reaches $\pmb{\rho}_c$ (i.e., $\pmb{\rho}(t) = \pmb{\rho}_c$ at some time $t$), the state has to stay at $\pmb{\rho}_c$ forever \cite{explanation01}. Therefore, upon increasing $t_f$, we expect $\pmb{\rho}(t)$ asymptotes to, but never reaches, $\pmb{\rho}_c$ during the singular control. By doing so, $\langle \pmb{\lambda} | \mathbf{f} (\pmb{\rho}) \rangle$ comes closer and closer to zero but never reaches zero.

%The solution we found is that as $t_f$ increases, $\pmb{\rho}(t)$ goes closer and closer to $\pmb{\rho}_c$ but never reaches $\pmb{\rho}_c$; by doing so, $\langle \pmb{\lambda} | \mathbf{f} (\pmb{\rho}) \rangle$ comes closer and closer to zero but never reaches zero. Because Eq.~\eqref{eqn:optimal_tf_condition} corresponds to the optimal $t_f$, we get an increasing target-state overlap $\langle \pmb{\rho}_f , \pmb{\rho} (t_f) \rangle $ as $t_f$ increases. 

\begin{figure}[ht]
\begin{center}
\includegraphics[width=0.6\textwidth]{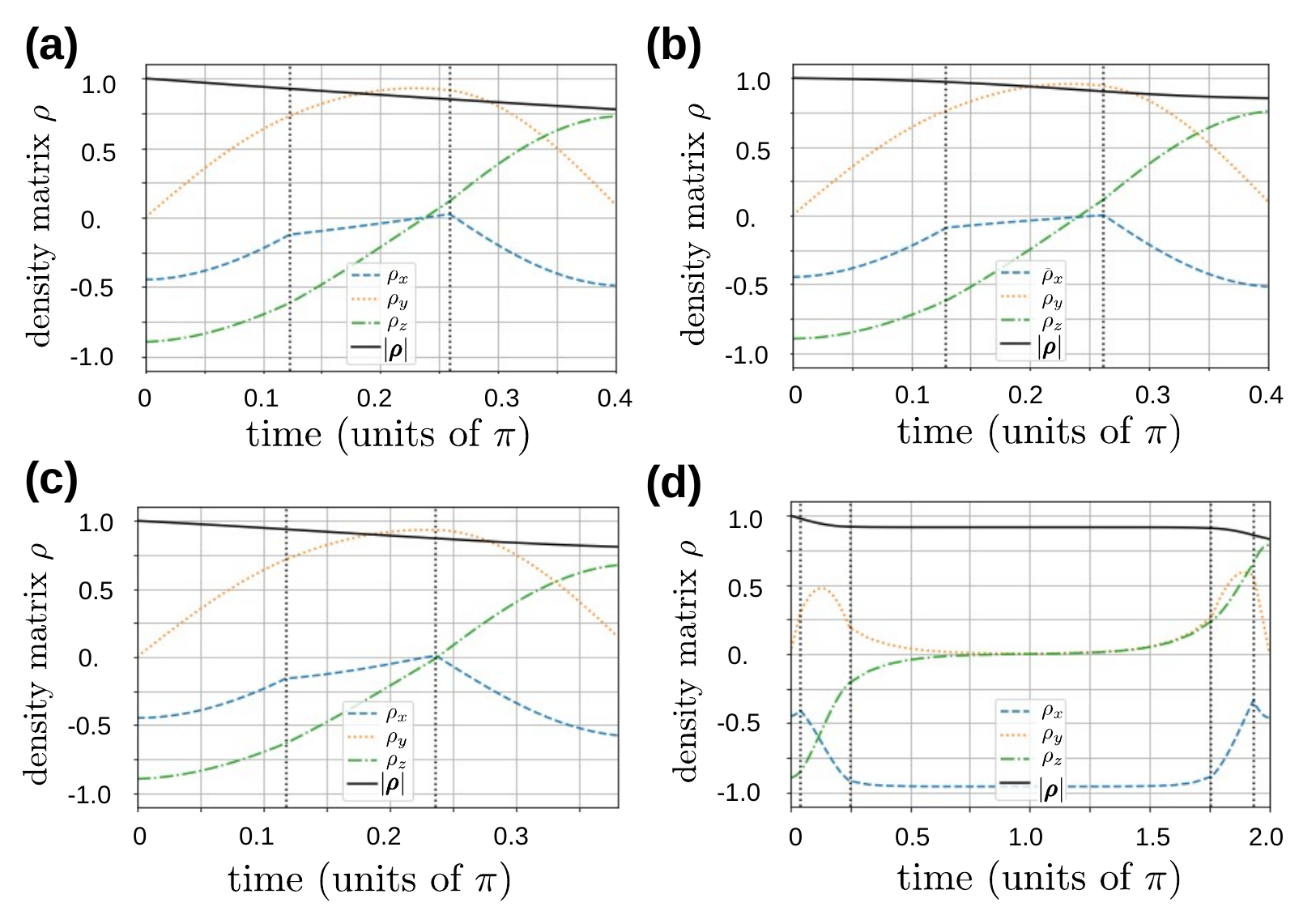}
\caption{ 
$\pmb{\rho}(t)$ during the optimal control for different dissipation channels. (a) uniform dissipation with $t_f=0.4 \pi$; (b) $\sigma_z$ dissipation channel with $t_f=0.4 \pi$. The trajectories are very similar for these cases. (c) and (d) $\sigma_x$ dissipation channel for (c) $t_f=0.38 \pi$  and (d) $t_f = 2.00 \pi$. At $t_f=0.38 \pi$ the trajectory is still similar to those in (a) and (b). As $t_f$ increases, $\pmb{\rho}(t)$ approaches the point $\pmb{\rho}_c = (\rho_x, 0, 0)$, where $u \rightarrow 0$ is an admissible control that leaves  $\pmb{\rho}_c$ unchanged. During the singular control in $\sigma_x$ dissipation channel, the amplitude of $|\pmb{\rho}(t) |$ decays slowly because of the small $\rho_y$ and $\rho_z$ components. Vertical lines indicate the switching times.
}
\label{fig:LZ_DM_tscan_DampX_Gamma0p1_xxp2_rhot}
 \end{center}
\end{figure}

For state preparation problems, two scenarios can occur: 
\bi 
\item Case (i): $\pmb{\rho}_c$ is approached when the terminal cost function (negative of target-state overlap)  decreases upon increasing $t_f$.
\item Case (ii): $\pmb{\rho}_c$ is approached when the terminal cost function increases upon increasing $t_f$. 
\ei
%We remind that the terminal cost function is the negative of the target-state overlap [Eq.~\eqref{eqn:cost_overlap}]. 
Because $\pmb{\rho}_c$ can never be reached, the target-state overlap in Case (i) will keep on increasing as $t_f$ increases; the optimal $t_f$ to maximize the target-state overlap is therefore {\em infinite} in this case. For Case (ii), the optimal $t_f$ to maximize the target-state overlap is finite; upon increasing $t_f$, the target-state overlap decays to a value larger than 0 (the value obtained by $\pmb{\rho}=0$).
Both cases are found in the numerical simulations. We emphasize that with the $\sigma_x$ dissipation channel, the optimal control {\em always} prevents the system from decaying to the maximum-entropy $\pmb{\rho}=0$ state at $t_f \rightarrow \infty$. 

A representative example of Case (i) is illustrated in Fig.~\ref{fig:LZ_DM_tscan_DampX_Gamma0p1_xxp2}(a), where the target-state overlap and the corresponding optimal control protocols using $\xi=0.2$ and $\Gamma_x = 0.1$ are plotted: upon increasing the evolution time $t_f$, the optimal protocol goes from XY to XSY to XSXY to XYSXY. In Fig.~\ref{fig:LZ_DM_tscan_DampX_Gamma0p1_xxp2} (b1)-(b4) we show that the optimality conditions are satisfied for the representative $t_f$ of each protocol. %As expected from the general considerations,
The most noticeable feature is the absence of a finite optimal $t_f$ --  $\langle \pmb{\rho}_{f},  \pmb{\rho} (t_{f}) \rangle$ keeps on increasing but saturates at a value smaller than one (about 0.91) as $t_f$ increases. 
%This is consistent with the conclusion drawn from general consideration, and we now analyze this example in more detail.  
%The non-existence of a finite optimal $t_f$ is indeed the key feature of Case (i) according to general considerations.
%implies that at some point of evolution, $\pmb{\rho}(t)$ has to approach a special state $\pmb{\rho}_c$ which is immune to the dissipation and there exists an admissible control that leaves $\pmb{\rho}_c$ unchanged. As the $\sigma_x$ dissipation channel only preserves the $\rho_x$ component, $\pmb{\rho}_c$ must be $(\rho_x, 0, 0)$ where the $\sigma_x$ dissipation channel has no effect. Eq.~\eqref{eqn:singular_3d} implies that the singular control at $\pmb{\rho}_c$ is zero, and $u=0$ control leaves $\pmb{\rho}_c$ unchanged. We therefore expect that time-optimal $\pmb{\rho}(t)$ stays close to $\pmb{\rho}_c$ (but never reaches $\pmb{\rho}_c$) for a long time when $t_f$ is large.
To examine this example in more detail, Fig.~\ref{fig:LZ_DM_tscan_DampX_Gamma0p1_xxp2_rhot} provides optimal trajectories of $\pmb{\rho}(t)$ for different dissipation channels. For the uniform, $\sigma_y$ (not shown), and $\sigma_z$ dissipation channels, the trajectories are very similar. In particular, the $\rho_x$ component first goes through zero and then approaches the target value [Fig.~\ref{fig:LZ_DM_tscan_DampX_Gamma0p1_xxp2_rhot}(a) and (b)]. For the $\sigma_x$ dissipation channel, the small-$t_f$ behavior is similar to those of other dissipation channels [see $t_f=0.38 \pi$ in Fig.~\ref{fig:LZ_DM_tscan_DampX_Gamma0p1_xxp2_rhot}(c)]. When $t_f$ increases, the time spent on the singular control increases correspondingly and the trajectory gets closer to $\pmb{\rho}_c$ during the singular control. Consequently the reduction of $|\pmb{\rho}|$ becomes extremely weak [see the solid curve in Fig.~\ref{fig:LZ_DM_tscan_DampX_Gamma0p1_xxp2_rhot}(d)] because most of $\pmb{\rho}(t)$ lies in its $\rho_x$ component.  In the example of $t_f=2.0 \pi$ shown in Fig.~\ref{fig:LZ_DM_tscan_DampX_Gamma0p1_xxp2_rhot}(d), $\pmb{\rho}(t)$ stays around $(\rho_x, 0, 0)$ between $t \sim \pi/2$ and $t \sim 3\pi/2$ to minimize the damping effect.

\begin{figure}[ht]
\begin{center}
\includegraphics[width=0.7\textwidth]{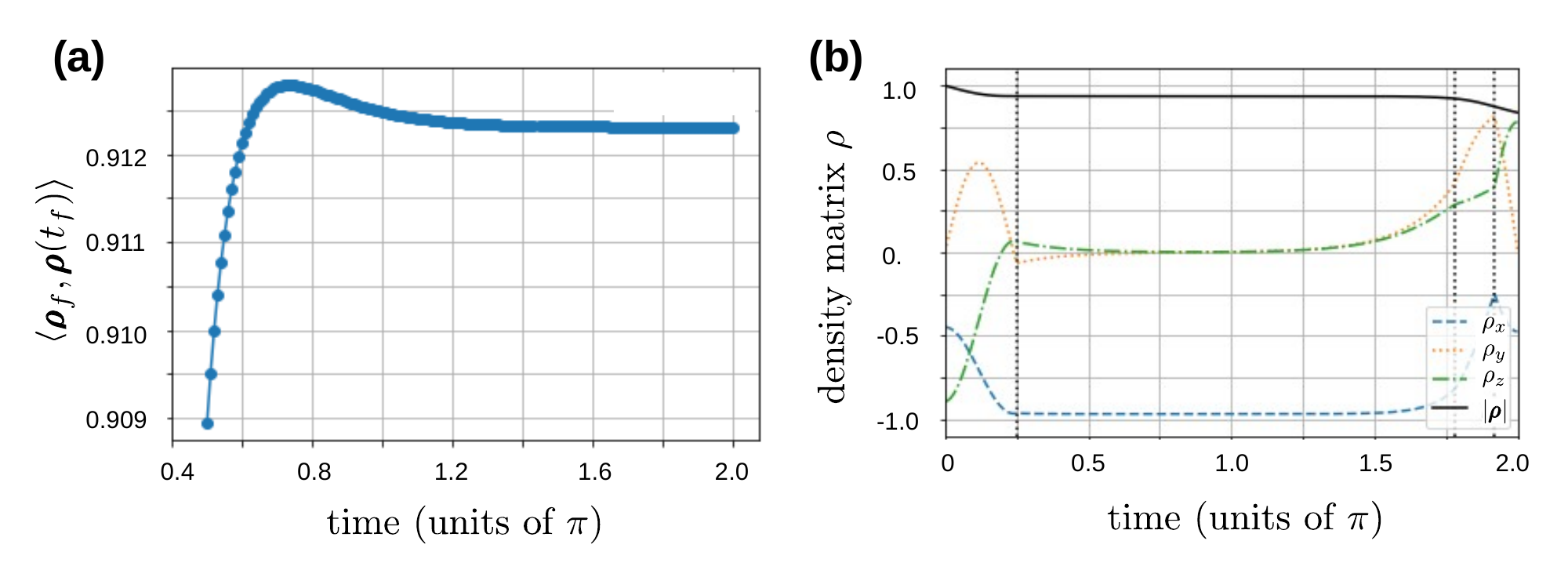}
\caption{ (a) The target-state overlap $\langle \pmb{\rho}_f, \pmb{\rho}(t_f) \rangle$ as a function of evolution time $t_f$ using $\xi=0.8$. The optimal control structure for $t_f \geq 0.5 \pi$ is found to be YSXY. 
An optimal $t_f$ is around 0.73 $\pi$. As $t_f$ increases, $\langle \pmb{\rho}_f, \pmb{\rho}(t_f) \rangle$ does not decay to zero. (b) The $\pmb{\rho}(t)$ for $t_f = 2.0 \pi$. $\pmb{\rho}(t)$ spends a almost 50\% of time close to $\pmb{\rho}_c = (\rho_x,0,0)$, the point where the dissipation has no effect. Three vertical dotted lines indicate the switching times when the control protocol changes from Y to S, from S to X, and from X to Y.  
}
\label{fig:sigma_x_xi=0.8_longtime}
 \end{center}
\end{figure} 

A representative example of Case (ii) is illustrated in Fig.~\ref{fig:sigma_x_xi=0.8_longtime}(a), where the target-state overlap for $0.5 \pi < t_f < 2.0 \pi$ using $\xi=0.8$ and $\Gamma_x = 0.1$ is plotted; the optimal protocol for $t_f \geq 0.5 \pi$ is YSXY. 
In this case, the optimal time that maximizes the target-state overlap is around $t=0.73 \pi$. Unlike other dissipation channels, $\langle \pmb{\rho}_f, \pmb{\rho}(t_f) \rangle$  does not decay to zero but approaches to a value about 0.91, which is the key feature of Case (ii). % according to general considerations. 
Fig.~\ref{fig:sigma_x_xi=0.8_longtime}(b) provides optimal trajectory of $\pmb{\rho}(t)$ for $t_f=2.0 \pi$. We again observe that  $\pmb{\rho}(t)$ stays around $(\rho_x, 0, 0)$ between $t \sim \pi/2$ and $t \sim 3\pi/2$ to minimize the damping effect. 
%\textcolor{blue}{
Our simulations indicate that when $|\xi| \lesssim 0.6$ and $\Gamma_x=0.1$, the $t_f \rightarrow \infty$ behavior is of Case (i) type; when $|\xi| > 0.6$ and $\Gamma_x=0.1$, the $t_f \rightarrow \infty$ behavior is of Case (ii) type. The transition between these two behaviors is determined by the relative position between the singular arc and the chosen initial and final states. In any case the system never decays to the maximum-entropy $\pmb{\rho}=0$ state. 

In fact, when the evolution time $t_f$ is sufficiently long, the system is steered to stay close to $\pmb{\rho}_c$ to minimize the decoherence effect. This $t_f \rightarrow \infty$ behavior appears to be independent of choices of initial and target states as far as the dimension of $\pmb{\rho}_c$ is not zero. We have performed several simulations using different states or using different Hamiltonians/dissipation channel (leading to a different $\pmb{\rho}_c$, not shown) to numerically verify this general behavior. As an illustration 
in Appendix \ref{subsec:sigma_x_memory} we provide an interesting case where the initial and target states are identical. It is somehow remarkable that the optimal control finds a path that can partially preserve the coherence even for the infinite evolution time. 
%}

%%%%%%%%%%%%%%%%%%

%%%%%%%%%%%%%%%%%%%%%%%%%
\section{Conclusion}

%SPeculate that $\pmb{\rho}_c$ is always reacheable?

We have applied Pontryagin's maximum principle to determine the time-optimal control that steers a general initial state to a general target state in a dissipative single qubit. The Hamiltonian is of Landau-Zener type, and we considered various loss channels described by a Lindblad master equation. Generally, the optimal protocol for time-optimal control problems is expected to have the bang-bang structure. 
However, at sufficiently long times we have found that all optimal control protocols, dissipative or not, include singular arcs. 
Determination of the time-dependent singular control is not straightforward. Using the geometric control technique we were able to obtain the allowed singular control without performing time integration, and obtained optimal protocols can be verified by the optimality conditions imposed by PMP.  

%We are able to 
%Without dissipation, the time-optimal control changes from bang-bang to bang-singular-bang as the evolution time increases.  
%The first class of problems considered is the quantum state preparation where the initial and the target state are different. 
With a generic dissipation, the target state can never be reached and there exists an optimal evolution time beyond which the dissipation prevents the system from getting closer to the target state. 
%\textcolor{red}{ 
For the $\sigma_x$ dissipation channel with a sufficiently long evolution time, however, the optimal control is found to take the qubit arbitrarily close to the decoherence free subspace during the evolution. This surprising feature can be traced to the presence of a one-dimensional null-space of the drift field. As a consequence, the target-state overlap always saturates at a value larger than zero as $t_f \rightarrow \infty$.  Depending on the relative positions between the singular arc and the chosen initial and final states, the optimal evolution time to maximize the target-state overlap can become {\em infinite}.  
For other dissipation channels where the null-space of the drift field has zero dimension, the optimal $t_f$ is always finite as the state will become maximally mixed at long times. If a qubit or two-level system has a dominant dissipation channel, our calculations indicate that the dissipation effect can be minimized by properly choosing the drift field. 
%}

%The second class of problems considered is the quantum state retention or quantum memory application where the initial and the final state are identical. In this case the presence of the dissipation demands an optimal evolution time as the density matrix will eventually lose all its coherence and decay to the maximum-entropy state. 
%%Finally the application of using qubit as a memory unit is exploited by setting the target state to be the initial state. PMP is then applied to find the optimal protocol. %The optimal protocol depends on the evolution time $t_f$. 
%The universal feature found by PMP is that the target-state overlap (which is also the initial-state expectation value) does not decay monotonously as a function of the evolution time -- although the dissipation always reduces the amplitude of $\pmb{\rho}$, there exists a time window where the final state becomes closer to the initial state upon increasing $t_f$. This non-monotonous dependence can be understood from the unitary dynamics where a state will always go back to itself after a certain amount of time. For the $\sigma_x$ dissipation channel, we find that the density matrix does not decay to the maximum-entropy state in the $t_f \rightarrow \infty$ limit and the underlying reason is also the one-dimensional null-space of the drift field. 
%

\section*{Acknowledgment}
%We thank ...... for very helpful discussions.
C.L. thanks Yanting Ma and Arvind Raghunathan (Mitsubishi Electric Research Laboratories) for very helpful discussions.
D.S. acknowledges support from the FWO as post-doctoral fellow of the Research Foundation -- Flanders. Invaluable comments from two anonymous referees are gratefully appreciated.
%%%%%%%%%%%%%%%%%%%%%%%%%
\appendix 

%\section{Results of $\sigma_z$ dissipation channels \label{App:sigma_z}} 

%In this Appendix we provide the results for the $\sigma_z$ dissipation channel. The equation of motion is 
%\beq 
%\frac{\dd }{\dd t} \pmb{\rho}  = 2 \mathbf{h} \times \pmb{\rho} 
%- \Gamma_z \mathbf{P}_z \, \pmb{\rho}.
%\eeq 
%with $\Gamma_z = 0.1$. Fig.~\ref{fig:LZ_DM_tscan_DampZ_Gamma0p1_xxp2}(a) summarizes the optimal control protocol as a function of the evolution time: it goes from XY to XSY as $t_f$ increases. The necessary conditions for optimality is checked in Fig.~\ref{fig:LZ_DM_tscan_DampZ_Gamma0p1_xxp2}(b1)-(b4). The behavior is qualitatively the same as that of the uniform dissipation discussed in Section \ref{subsec:uniform_diss}. A minor quantitative difference is that the $\sigma_z$ dissipation channel has a larger target-state overlap (or smaller cost function) due to its weaker damping effect.

%%%

\section{Optimal control using gradient-based algorithm and switching function \label{subsec:numeric}}

\begin{figure}[ht]
\begin{center}
\includegraphics[width=0.4\textwidth]{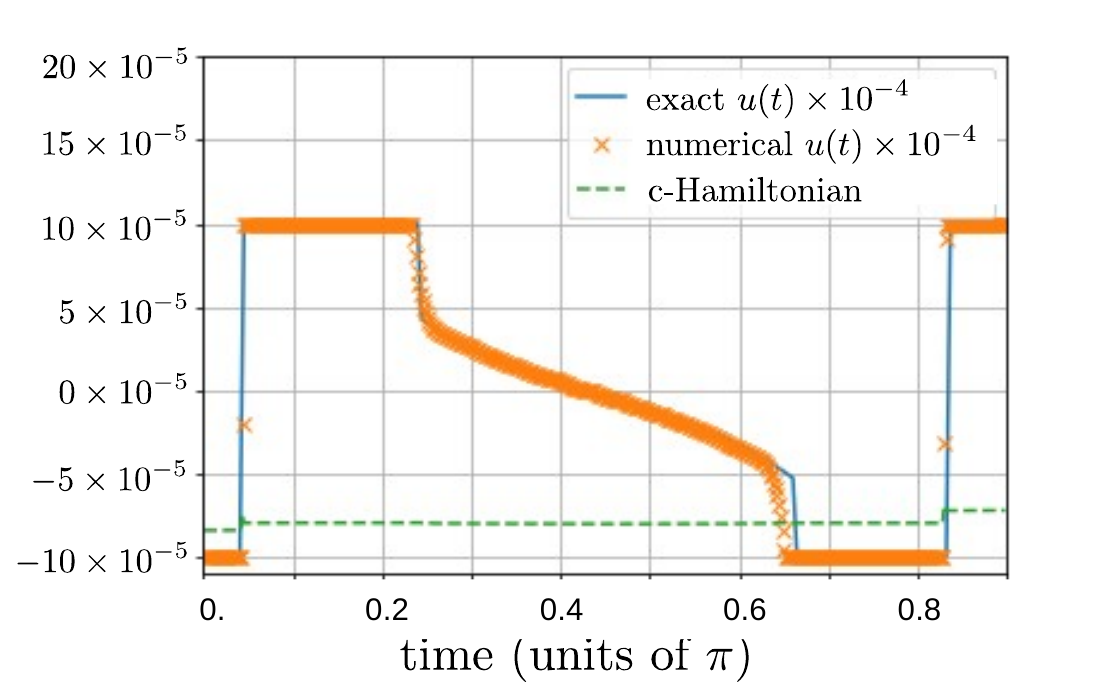}
\caption{ The optimal control obtained from the gradient-based numerical optimization (cross symbol, 500 points) and the numerically exact procedure (solid). A good agreement is seen, especially for the singular part. The parameters are the same as those used in Fig \ref{fig:LZ_DM_tscan_DampX_Gamma0p1_xxp2}, and the exact optimal control is taken from Fig.\ref{fig:LZ_DM_tscan_DampX_Gamma0p1_xxp2} (b4). The c-Hamiltonian obtained from numerical optimization is also given. There is a small discontinuity across each switching time because the exact switching times are not captured.  
}
\label{fig:tf_0p9pi_numerics}
\end{center}
\end{figure} 

The terminal cost function can be directly minimized by discretizing $u(t)$ as $u(t_1), u(t_2), ..., u(t_N)$; the optimal control corresponds to $\{ u(t_i) \}$ that minimizes the terminal cost function.
Using the switching function as the gradient, the optimal control can be numerically obtained by iterating
\begin{equation}
\begin{aligned}
 & u^{(n+1)}(t_i) =  u^{(n)}(t_i) - \lambda \Phi(t_i) %\\
% & \text{forcing } ,
\end{aligned}
\label{eqn:numerical_updating}
\end{equation}
with $|u^{(n+1)}(t_i) |\leq 1$, until a stopping criterion is satisfied. Here $\lambda>0$ is the updating rate; if $|u^{(n+1)}(t_i)|>1$, $u^{(n+1)}(t_i)$ is chosen to be the closest extreme value.  
%When the optimal control involves a singular control where $\Phi(t)=0$ over a finite interval of time, one might concern that the gradient based method does not provide sufficiently significant update. Our numerical simulations show that this is {\em not} the case, as $\Phi(t)=0$ only happens at the optimal solution. 

An example of $\sigma_x$ dissipation channel with $\Gamma_x = 0.2$, $\xi=0.1$, and $t_f=0.9\pi$ is provided in Fig.~\ref{fig:tf_0p9pi_numerics}.
This is the most complicated case in the dissipative qubit.  In this simulation, the terminal cost function is chosen to be $\mathcal{C}(t_f) = \sum_{ij} | \rho_{f, ij} - \rho(t_f)_{ij} |^2$ with $\rho_f = | \psi_f \rangle \langle \psi_f |$. 
As shown in Fig.\ref{fig:LZ_DM_tscan_DampX_Gamma0p1_xxp2} (b4), the optimal control has an XYSXY structure.  We discretize $u(t)$ into 500 points and use conjugate gradient optimization method to get the optimal control. The  result is very close to the numerical exact solution obtained in Fig.\ref{fig:LZ_DM_tscan_DampX_Gamma0p1_xxp2} (b4) and is not sensitive to the initial guess. The corresponding c-Hamiltonian is also plotted (dashed curve) in Fig.~\ref{fig:tf_0p9pi_numerics}; because of the different terminal cost function, this value is different from the c-Hamiltonian of Fig.\ref{fig:LZ_DM_tscan_DampX_Gamma0p1_xxp2} (b4). 
Although close, $\mathcal{H}_c$ is not exactly a constant over the whole evolution time. In particular, there is a small jump across each switching time because the exact switching times are not captured. When the exact solution is not known (such as problems of higher dimension), the optimality conditions listed in Eqs.\eqref{eqn:necc_cond_state_variables} can be served to quantify the quality of any numerically solution. 

\section{Optimal protocol for the state retention under the $\sigma_x$ dissipation channel \label{subsec:sigma_x_memory}} 

\begin{figure}[ht]
\begin{center}
\includegraphics[width=0.7\textwidth]{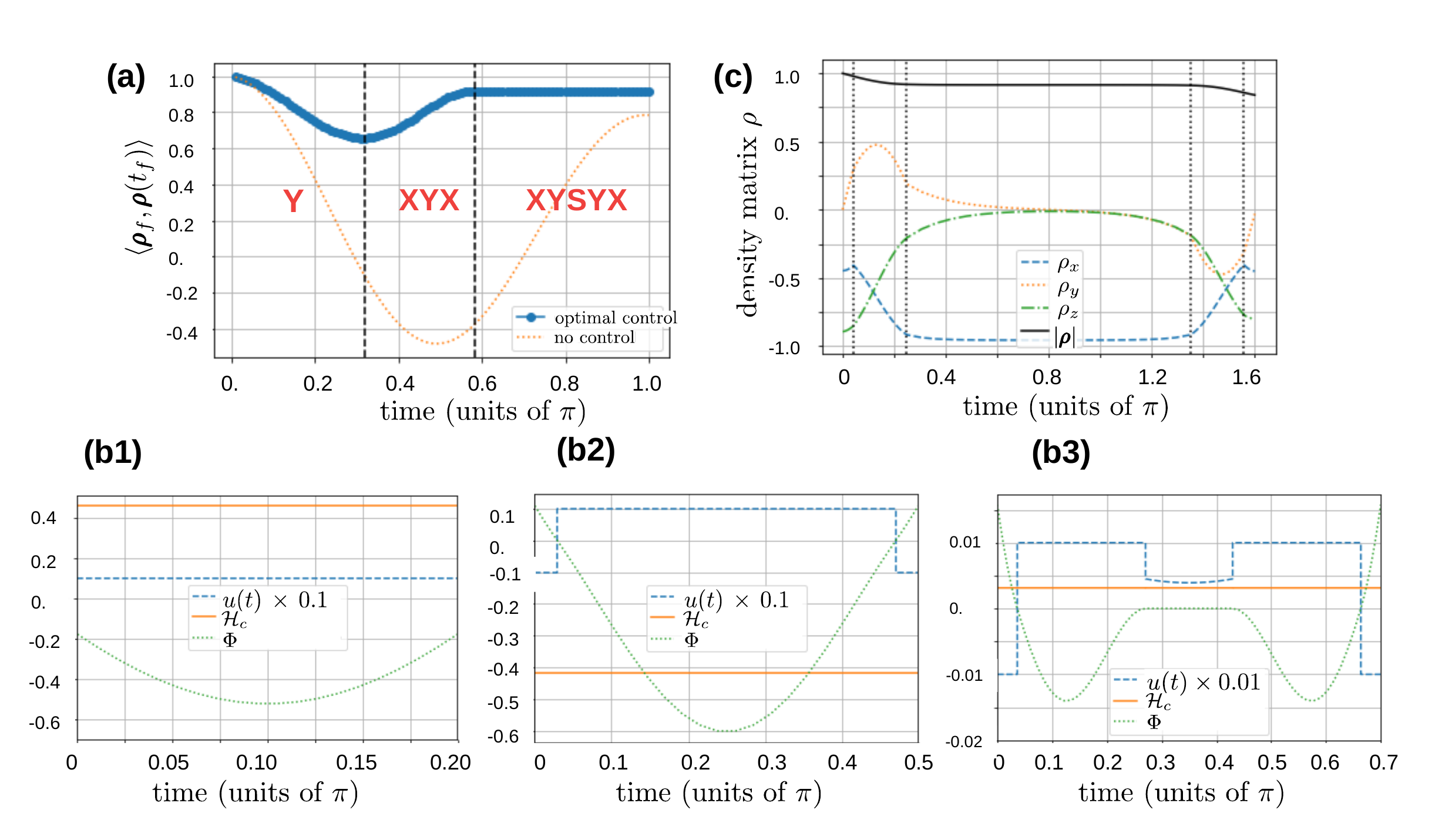}
\caption{ The quantum state retention with the $\sigma_x$ dissipation channel. (a) The target-state overlap $\langle \pmb{\rho}_f, \pmb{\rho}(t_f) \rangle$ as a function of evolution time $t_f$. The solid-circle curve is obtained using optimal control; the dotted curve is obtained without any control ($u(t)=0$). The target-state overlap obtained using optimal control is larger than that obtained with no control. Vertical dashed (black) lines are boundaries of different optimal control structures; corresponding optimal controls are indicated in red. A local maximum occurs at $t_f \approx 0.58 \pi$ for the optimal control; at $t_f \approx \pi$ for zero control. As $t_f$ increases, $\langle \pmb{\rho}_f , \pmb{\rho}(t_f) \rangle$ does not decay to zero.  (b1)-(b3) The optimal control for three different evolution times: (b1) $t_f=0.2 \pi$; (b2) $t_f=0.5 \pi$; (b3) $t_f=0.7 \pi$. Dashed curves: scaled control; solid curves: c-Hamiltonian; dotted curves: switching function. As the evolution time increases, the optimal control changes from Y (b1) to XYX (b2) to XYSYX (b3). (c) $\pmb{\rho}(t)$ for $t_f = 1.6\pi$. Around $t=0.8 \pi$, $\pmb{\rho}$ is close to $\pmb{\rho}_c = (\rho_x,0,0)$. 
}
\label{fig:LZ_Memory_tscan_DampX_Gamma0p1_xxp2}
\end{center}
\end{figure}

%(a) The target-state overlap $\langle \pmb{\rho}_f , \pmb{\rho} (t_f) \rangle $ as a function of evolution time $t_f$. Vertical dashed (black) lines are boundaries of different optimal control structures; corresponding optimal controls are indicated in red. (b1)-(b4) The optimal controls at four representative computational times(b1)-(b4) The optimal controls at four representative computational times: (b1) $t_f=0.2 \pi$; (b2) $t_f=0.35 \pi$; (b3) $t_f=0.419 \pi$ (optimal $t_f$); (b4) $t_f=0.90 \pi$. As the evolution time increases, the optimal control changes from XY (b1) to XSY (b2) to XSXY (b3) to XYSXY (b4).

As an interesting generalization, we consider the state retention problem where the target state is the initial state [$| \psi_i \rangle = | \psi_f \rangle = $ the first equation of Eq.~\eqref{eqn:init_final}] and determine the optimal protocol for the $\sigma_x$ dissipation channel.  
Fig.~\ref{fig:LZ_Memory_tscan_DampX_Gamma0p1_xxp2} summarizes the optimal control for  $\xi=0.2$, $\Gamma_x = 0.1$. As given in Fig.~\ref{fig:LZ_Memory_tscan_DampX_Gamma0p1_xxp2}(a), the optimal protocol changes from Y to XYX to XYSYX as the evolution time increases. 
%obtained using optimal control is larger than that obtained with no control. 
The target-state overlap $\langle \pmb{\rho}_f, \pmb{\rho}(t_f) \rangle  = \langle \pmb{\rho}_i, \pmb{\rho}(t_f) \rangle$  displays a global maximum at $t_f=0$, a local maximum around $t_f=0.58 \pi$, and asymptotes to about 0.92 as $t_f \rightarrow \infty$. Because $| \psi_i \rangle$ and $| \psi_f \rangle$, $\langle \pmb{\rho}_i, \pmb{\rho}(t_f=0) \rangle = \langle \pmb{\rho}_i, \pmb{\rho}_i \rangle = 1$ is automatically the global maximum.
The second local maximum can be understood from the unitary dynamics where a state will always go back to itself after a certain amount of time. The asymptotic behavior corresponds to Case (ii) scenario discussed in Section \ref{subsec:sigma_x}. At any $t_f$, the target-state overlap using optimal control is larger than that with zero control. 
The necessary conditions are checked and the representative control protocols are shown in Fig.~\ref{fig:LZ_Memory_tscan_DampX_Gamma0p1_xxp2}(b1)-(b3). In Fig.~\ref{fig:LZ_Memory_tscan_DampX_Gamma0p1_xxp2}(c) the optimal trajectory of $\pmb{\rho}(t)$ for $t_f=1.6 \pi$. We see that 
the amplitude of $\pmb{\rho}$ almost remains unchanged during the singular control as the quantum state spends most of time around $\pmb{\rho}_c = (\rho_x,0,0)$ where the $\sigma_x$ dissipation channel has no effect. The analytical analysis provided in Section \ref{subsec:sigma_x} does not rule out the possibility that the local maximum appears at $t_f \rightarrow \infty$ (i.e., no finite local maximum), but we do not find it between $\xi=-1$ to $1$. 
%%%%%%%%%%%%%%%%%%%%%%%%%%%%%%%%%%%%%%%%%%%
\bibliography{QC_optimal_control}

\begin{thebibliography}{55}
\bibitem[{exp({\natexlab{a}})}]{explanation00}
\bibinfo{note}{This is derived as follows. From Eq.~\eqref{eqn:vector_pauli_x}
  $V_x$ reduces to $-2 \partial_\theta$ at $\phi=\pi/2$, so two dynamical
  variables satisfy $\dot{\theta}=-2$ and $\dot{\phi}=0$. These equations imply
  that the time to move from $\theta_0$ to $\theta_1$ at a given $\phi$ is
  $|\theta_0 - \theta_1|/2$.}
\bibitem[{exp({\natexlab{b}})}]{explanation01}
\bibinfo{note}{This statement uses both (i) the singular arc guarantees
  $\langle \pmb{\lambda} | \mathbf{g} \rangle =0$ and (ii) the optimal $t_f$
  requires $\mathcal{H}_c = \langle \pmb{\lambda} | \mathbf{f} \rangle + u(t)
  \langle \pmb{\lambda} | \mathbf{g} \rangle = 0$. During the singular control,
  $ \mathcal{H}_c = \langle \pmb{\lambda} | \mathbf{f} \rangle$. If $
  \mathcal{H}_c=0$ is satisfied by the vanishing vector field $\mathbf{f}(
  \pmb{\rho}_c )$ (not by the vanishing inner product), then $\pmb{\rho}(t)$
  has to stay in the subspace defined by $\pmb{\rho}_c$; otherwise
  $\mathcal{H}_c$ becomes non-zero and the trajectory cannot be an
  optimal-$t_f$ solution.}
\end{thebibliography}
%\bibliographystyle{unsrt}
%\begin{thebibliography}{10}
%\end{thebibliography}

%\appendix

\end{document}